\documentclass[a4paper,fleqn,usenatbib]{mnras}

\usepackage{newtxtext,newtxmath}
\usepackage[T1]{fontenc}
\usepackage{ae,aecompl}

\usepackage{graphicx}	
\usepackage{amsmath}	
\usepackage{amssymb}	

\usepackage{booktabs}      
\usepackage{array}            

\newcommand{\gsim}{~\rlap{$>$}{\lower 1.0ex\hbox{$\sim$}}}

\newcommand{\lsim}{~\rlap{$<$}{\lower 1.0ex\hbox{$\sim$}}}

\newcommand{\Meraxes}{\textsc{Meraxes}}

\newcommand{\tocf}{\textsc{21cmfast}}

\newcommand{\ra}[1]{\renewcommand{\arraystretch}{#1}}

\title[DRAGONS: Bubbles at dawn]{Dark-ages~reionization~\&~galaxy~formation~simulation~XII: Bubbles at dawn}

\author[P. M. Geil et al.]{
\parbox{\textwidth}{
Paul M. Geil$^{1,2}$\thanks{E-mail: geil.p@unimelb.edu.au (PMG)},
Simon J. Mutch$^{1}$,
Gregory B. Poole$^{1}$,
Alan R. Duffy$^{3}$,
Andrei Mesinger$^{4}$
and J. Stuart B. Wyithe$^{1,2}$}\vspace{0.2cm}
\\
$^{1}$School of Physics, The University of Melbourne, Parkville, Victoria 3010, Australia\\
$^{2}$ARC Centre of Excellence for All-sky Astrophysics (CAASTRO) \\
$^{3}$Centre for Astrophysics and Supercomputing, Swinburne University of Technology, PO Box 218, Hawthorn, VIC 3122, Australia\\
$^{4}$Scuola Normale Superiore, Piazza dei Cavalieri 7, I-56126 Pisa, Italy
}

\date{Accepted XXX. Received YYY; in original form ZZZ}

\pubyear{2017}

\begin{document}
\label{firstpage}
\pagerange{\pageref{firstpage}--\pageref{lastpage}}
\maketitle

\begin{abstract}
Direct detection of regions of ionized hydrogen (H\,\textsc{ii}) has been suggested as a promising probe of cosmic reionization. Observing the redshifted 21-cm signal of hydrogen from the epoch of reionization (EoR) is a key scientific driver behind new-generation, low-frequency radio interferometers. We investigate the feasibility of combining low-frequency observations with the Square Kilometre Array and near infra-red survey data of the \emph{Wide-Field Infrared Survey Telescope} to detect cosmic reionization by imaging H\,\textsc{ii} bubbles surrounding massive galaxies during the cosmic dawn. While individual bubbles will be too small to be detected, we find that by stacking redshifted 21-cm spectra centred on known galaxies, it will be possible to directly detect the EoR at $z \sim 9$--12, and to place qualitative constraints on the evolution of the spin temperature of the intergalactic medium (IGM) at $z \gsim ~ 9$. In particular, given a detection of ionized bubbles using this technique, it is possible to determine if the IGM surrounding them is typically in absorption or emission. Determining the globally-averaged neutral fraction of the IGM using this method will prove more difficult due to degeneracy with the average size of H\,\textsc{ii} regions.
\end{abstract}



\section{Introduction}
\label{sec:Introduction}

Cosmic hydrogen is believed to have been reionized by ultraviolet (UV) radiation produced by stars and quasars. The period from the formation of the first ionizing sources to when the intergalactic medium (IGM) was completely ionized is commonly known as the epoch of reionization (EoR). However, due to formidable challenges in observation and simulation, our knowledge of this process is lacking. Knowing how reionization occurred, both in time and space, would not only dramatically improve our understanding of the evolution and properties the IGM, but also the formation and role of the ionizing sources responsible during this period.

Observations of high-redshift sources and the cosmic microwave background (CMB) have allowed some constraints to be placed on the timing and duration of the EoR. For example, Gunn-Peterson absorption troughs in quasar Lyman-$\alpha$ spectra set a lower limit for the end of reionization at $z \sim 6$ \citep{Fan2006, Mortlock2011}. Additionally, CMB observations provide a measure of the total optical depth to electron scattering. Since this is an integrated quantity from the surface of last scattering ($z \sim 1100$), it cannot, on its own, distinguish between different reionization histories. However, depending on the model of reionization adopted, the average redshift at which reionization is half complete is found to lie between $z = 7.8$ and 8.8 \citep{PLANCK2016}. Recent analysis by \cite{Greig2016} implies that reionization is not yet complete by $z = 7.1$, with the volume-weighted IGM neutral fraction constrained to $0.40^{+0.41}_{-0.32}$ at 2$\sigma$.

A far more promising observational strategy to constrain reionization is to directly measure the emission from the 21-cm spin-flip transition of neutral hydrogen (H\,\textsc{i}). Due to cosmic expansion, the frequency of this radiation is now $< 200$\,MHz. Various experiments are underway or planned to measure the cosmic 21-cm signal as a function of frequency (and therefore redshift, time or distance) utilising different methods. One approach is to measure the spatially-averaged global signal using a single-dipole antenna, e.g., EDGES, \citet{Bowman2008}; DARE, \citet{DARE2012}; SARAS, \citet{SARAS2013}. Another is to measure the signal's spatial fluctuations interferometrically (using, e.g., LOFAR\footnote{\url{http://www.lofar.org}}, GMRT\footnote{\url{http://www.ncra.tifr.res.in/ncra/gmrt}}, PAPER\footnote{\url{http://eor.berkeley.edu}}, MWA\footnote{\url{http://www.mwatelescope.org}}, HERA\footnote{\url{http://reionization.org}}, SKA\footnote{\url{http://www.skatelescope.org}}). For some instruments the latter approach can yield both high-resolution tomographic images of the ionized structure and statistical measurements (such as the 21-cm power spectrum) allowing us to learn about the properties of the reionization sources and sinks in far greater detail than simple timing estimates. For reviews of the EoR science and 21-cm detection experiments, see e.g., \citet{MW2010} and \citet{Koopmans2015}.

In this work we use simulations to investigate structures of ionized hydrogen (H\,\textsc{ii}) surrounding the first galaxies during the early stages of the EoR ($z \gsim ~ 9$). During this era---known also as the cosmic dawn---these regions appear as isolated `bubbles'. We begin by discussing the ionized regions associated with simulation analogues of the highest-known redshift galaxy to date (GN-z11). We then move on to consider the wider population of bubbles in our simulation, establishing a relationship between their size, and redshift and luminosity of the brightest galaxy within them. We apply this simulation-based empirical relationship to explore the utility of an image regime-based EoR detection strategy that synergises the proposed \emph{Wide-Field Infrared Survey Telescope}'s (\emph{WFIRST}\footnote{\url{http://wfirst.gsfc.nasa.gov}}) High Latitude Survey (HLS) and deep integrations of the redshifted 21-cm signal using the planned low-frequency Square Kilometre Array (SKA1-LOW). Our direct detection stategy is similar to those proposed targeting regions of ionized hydrogen surrounding high-luminosity quasars \citep[see, e.g.,][]{Wyithe2005, Kohler2005, Geil2008} but is able to push the detection redshift beyond what is possible using quasars alone due to their relatively low population at $z > 8$. Other techniques for probing individual sources have been presented, both midway through the EoR and at very high redshift ($z \sim 15$), such as visibility-based methods using matched filtering \citep[e.g.][]{Datta2007, Datta2012, Majumdar2012, Ghara2016}, and using imaging \citep{Ghara2017}. Some of these works also assess the prospects of constraining properties of the high-redshift IGM (such as its globally-averaged neutral fraction) and the sources responsible for its reionization.

This paper is structured as follows. In Section~\ref{sec:The DRAGONS Simulation} we give a brief overview of the DRAGONS simulation used in this paper. Section~\ref{sec:HII regions surrounding the first galaxies} explains the motivation behind this work and presents our bubble size--galaxy redshift and luminosity relation. Section~\ref{sec:Detectability} describes our detection strategy and presents our detectability results. Section~\ref{sec:IGM properties} explores simple methods to constrain the spin temperature and globally-averaged ionization state of the high-redshift IGM. We address some additional details that may potentially impact our results in Section~\ref{sec:Discussion} before presenting a summary in Section~\ref{sec:Summary and conclusions}. We include an appendix containing supporting material detailing the model fitting, UV luminosity functions (UVLFs) and instrumental noise estimates used in this work. All globally-averaged quantities (e.g. neutral fraction) are volume weighted, and distances are given in comoving units unless stated otherwise. Absolute magnitudes used throughout are given in the AB system \citep{OG1983}, are intrinsic, and have been calculated using the methodology described in \citet{DRAGONS4}, assuming a standard \citet{Salpeter1955} initial mass function with upper and lower mass limits of 0.1\,M$_\odot$ and 120\,M$_\odot$, respectively. Our choice of cosmology is the standard spatially-flat \textit{Planck} $\Lambda$CDM cosmology \citep{PLANCK2015} $(h, \Omega_{\rm{m}}, \Omega_{\rm{b}}, \Omega_\Lambda, \sigma_8, n_{\rm{s}})$ = $(0.678, 0.308, 0.0484, 0.692, 0.815, 0.968)$.

\section{The DRAGONS Simulation}
\label{sec:The DRAGONS Simulation}

The Dark-ages, Reionization And Galaxy-formation Observables from Numerical Simulations (DRAGONS\footnote{\url{http://dragons.ph.unimelb.edu.au}}) project was specifically designed to study the formation of the first galaxies and cosmic reionization. It integrates a semi-numerical calculation of reionization (\tocf) within a semi-analytic model of galaxy formation (\Meraxes) built upon an $N$-body simulation (\emph{Tiamat}). This gives a self-consistantly coupled reionization model which accounts for feedback due to both supernovae~(SN) and the ionizing UV background from stars within galaxies. A unique feature of DRAGONS is that it utilises \emph{horizontal} rather than vertical dark matter halo merger trees. This allows it to correctly simulate how galaxies influence each others' evolution by way of their ionizing flux. \emph{Tiamat} has a sufficently large volume (cube of sides 100\,Mpc in length) to investigate cosmic evolution while still achieving a mass resolution approaching the atomic cooling mass threshold. \emph{Tiamat} also has excellent temporal resolution with 100 equally-spaced snapshots between redshifts 5--35, giving a cadence of about 11\,Myr. This means the stochastic effects of star formation and SN feedback on reionization are accurately captured. Complete descriptions of \emph{Tiamat} and \Meraxes~are given in \citet{DRAGONS1} (Paper-I) and \citet{DRAGONS3} (Paper-III), respectively, while details of the \tocf~algorithm are described in \citet{MFC2011}. \citet{DRAGONS5} (Paper-V) investigates the effect of galaxy-formation physics on the morphology and statistical signatures of reionization.

The \Meraxes~model used in this work is the fiducial model described in Papers-III and V. This model has been calibrated so as to reproduce the observed evolution of the galaxy stellar mass function from $z = 5$--7 (see Paper-III) and the latest \emph{Planck} optical depth measurements \citep{PLANCK2015}. All output fields (e.g. density, stellar mass and ionization fraction) have been regularly gridded over $512^3$ voxels.

\section{HII regions surrounding the first galaxies}
\label{sec:HII regions surrounding the first galaxies}

\subsection{GN-z11 analogues: DR-1 and DR-2}
\label{sec:GN-z11 analogues: DR-1 and DR-2}

Motivated by the identification of the surprisingly bright and massive galaxy GN-z11 at $z = 11.1$ by \citet{OESCH2016}, \citet{DRAGONS6} (Paper-VI) investigates the origin and fate of such objects using DRAGONS. Two analogue galaxies of similar luminosity and stellar mass to GN-z11, labelled DR-1 and DR-2, were found within the \emph{Tiamat} volume and show excellent agreement with all available observationally derived properties of this object. Maintaining this motivation, here we briefly summarise aspects of these objects' impact on the IGM in terms of reionization.

With the {\it Tiamat} volume gridded to $512^3$, the voxels containing DR-1 and DR-2 are first fully ionized at $z_{\rm ion} = 17.8$ and 17.1, respectively. While these objects were not the first sources to begin ionizing the IGM in our simulation (on scales corresponding to this grid resolution), they were among the first, with the majority of voxels being ionized after $z \approx 7.9$ (approximately corresponding to the redshift at which half the hydrogen in the IGM is ionized). This is not surprising given these objects' early star formation histories (see Paper-VI) and the fact that they are found in highly overdense regions.

The ionized bubbles surrounding DR-1 and DR-2 (and the other less massive and less luminous galaxies within them) at $z = 11.1$ are approximately spherical. This is due to the lack of overlapping bubbles surrounding other galaxies in their vicinity. At this redshift both DR-1 and DR-2 lie close to the centroid of their bubbles. We estimate individual bubble size using a three-dimensional ray tracing technique, centred upon the brightest galaxy in the bubble, which measures the distance to an ionization phase transition (demarked by a step to a voxel that is more than 50~per~cent neutral) in $\geq 10^3$ randomly chosen directions. In the case of DR-1 and DR-2 at $z = 11.1$ this provides an accurate and precise estimate of bubble radius. However, at later times (when there is overlap and the brightest galaxy in the bubble may be off-centre) the resulting sampled radius distribution has higher variance (which can be used to mark the approximate transition from an isolated bubble to an overlapping region). Using this method we find that the average diameters of the bubbles surrounding DR-1 and DR-2 at $z = 11.1$ are $\approx 10$ and 8\,Mpc, respectively. At this redshift the globally-averaged neutral fraction, $\bar{x}_{\rm H\textsc{i}}$, of our fiducially-modelled IGM is 0.976, hence these two bubbles alone (out of\,$\gsim ~ 600$) make up just under 3~per~cent of the total ionized volume.

\begin{figure*}
\includegraphics[width=\textwidth]{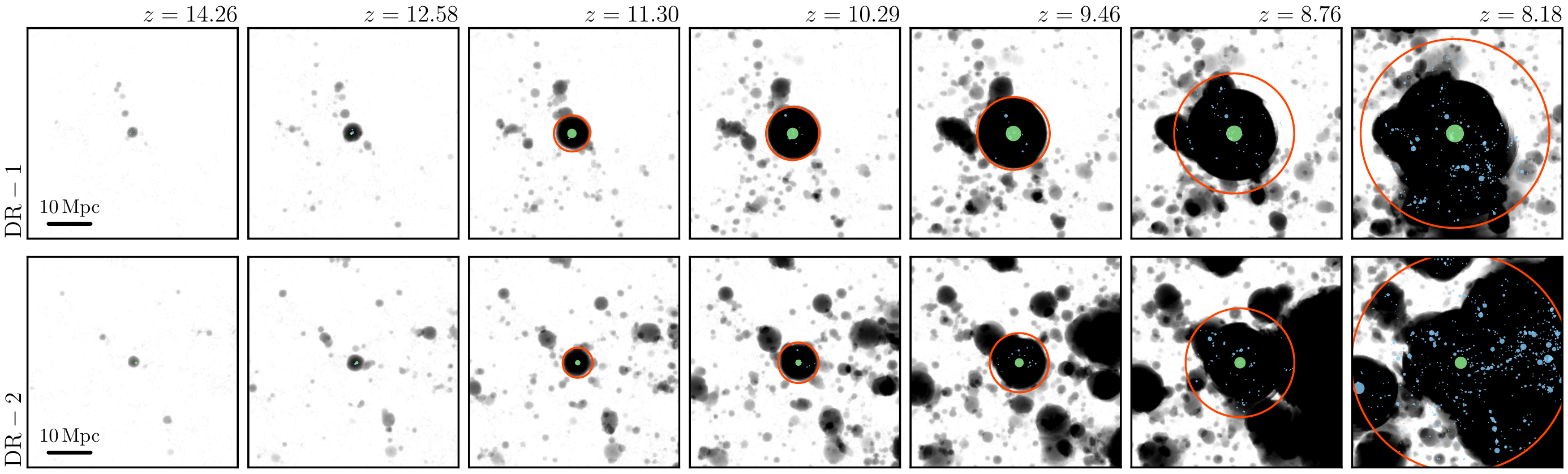}
\caption{Slices through the ionization fields surrounding DR-1 (top) and DR-2 (bottom) at selected redshifts showing the evolution of bubble size. The positions of DR-1 and DR-2 are indicated by the central (green) filled circles. The projected positions of other galaxies within the average radius of the bubble (shown by the red circle for panels with $z < 12$) are indictated by the other (blue) filled circles. The area of each galaxy's marker is proportional to its UV luminosity. Each slice is 5\,Mpc deep.}
\label{fig:F1-nf_seq}
\end{figure*}

Figure~\ref{fig:F1-nf_seq} shows zoomed-in slices through the ionization fields surrounding DR-1 and DR-2 at selected redshifts, showing the evolution of bubble size. The positions of DR-1 and DR-2 are indicated by the central (green) filled circles. The projected positions of other galaxies (as faint as $M_{\rm UV} = -17.25$) within the average radius of the bubble (shown by the red circle for panels with $z < 12$) are indictated by the other (blue) filled circles. The area of each galaxy's marker is proportional to its UV luminosity. By visual inspection these bubbles cease to be isolated regions, and are also driven by many less luminous galaxies, from $z \approx 9$.

\begin{figure}
\includegraphics[width = \columnwidth]{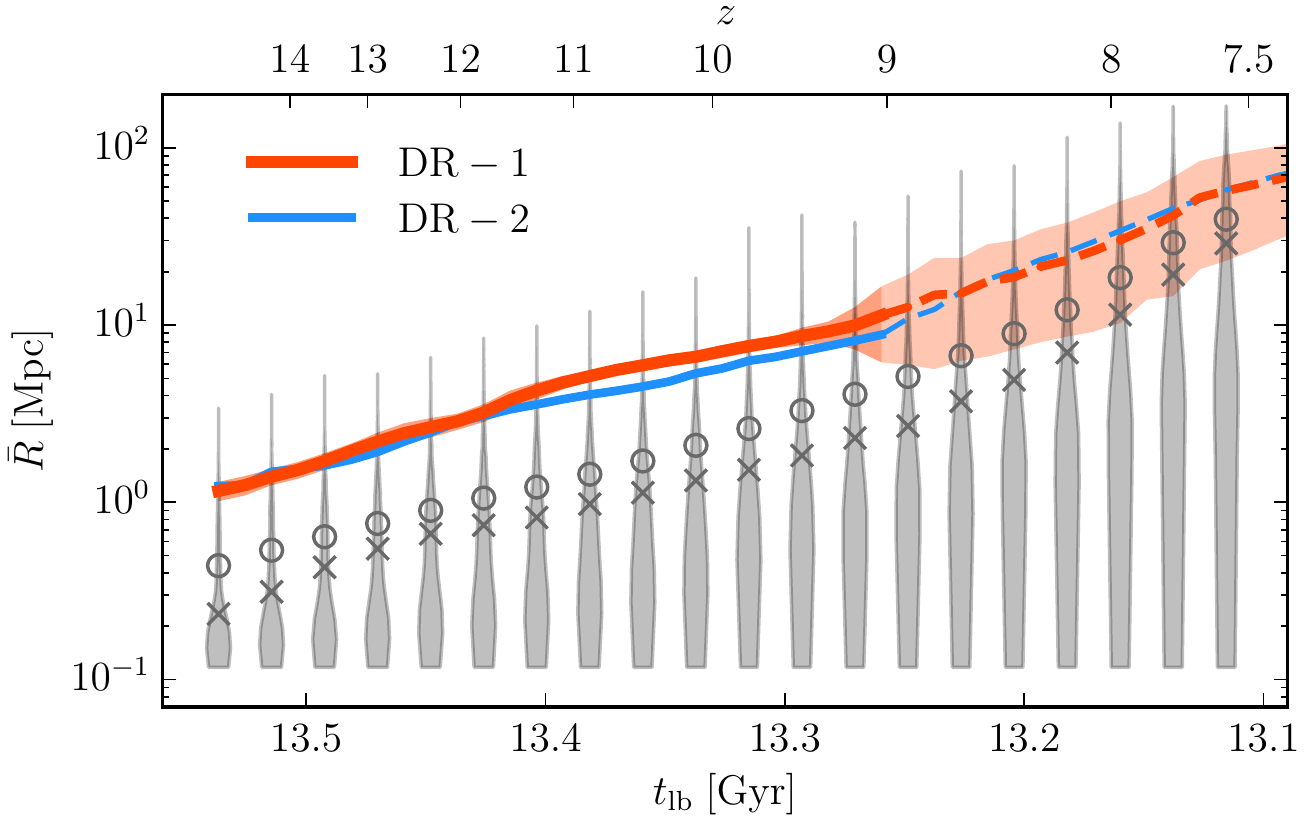}
\caption{Evolution of the size of bubbles surrounding DR-1 (red) and DR-2 (blue) as a function of redshift and lookback time. The red shaded region indicates the $\pm 1\sigma$ range in radius for DR-1, calculated using the ray tracing technique described in Section~\ref{sec:GN-z11 analogues: DR-1 and DR-2}. Dashed lines are used for when the error in radius is more than half the radius of the bubble (marking the approximate transition from an isolated bubble to an overlapping region). For comparison, the probability densities of the size of other ionized regions in the simulation at selected redshifts are shown by the violin plot (grey; with circles and crosses indicating their means and medians, respectively).}
\label{fig:F2-R_vs_lbt}
\end{figure}

In order to compare the bubbles surrounding DR-1 and DR-2 to others in the simulation, Figure~\ref{fig:F2-R_vs_lbt} shows where they lie in the size distribution of all ionized regions in the simulation as a function of redshift (and lookback time). The average size of the bubbles surrounding DR-1 and DR-2 are shown by the thick red and thinner blue lines, respectively. The shaded region indicates the $\pm 1\sigma$ range in radius for DR-1 (the uncertainty for DR-2 is not shown, but is similar to that of DR-1), calculated using the ray tracing technique described above. Dashed extensions of the lines show when the error in radius is more than half the radius of the bubble, marking the approximate transition from an isolated bubble to an overlapping region. For comparison, the sizes of other ionized regions in the simulation\footnote{We calculate the bubble size distribution for each snapshot using the Monte Carlo method described in \citet{MF2007}. This is the same as the ray tracing technique used to estimate individual bubble size described earlier, however, in this method, a voxel that is more than 50 per cent ionized is first randomly selected from the full gridded simulation volume and its distance from an ionization phase transition in a randomly chosen direction is recorded. This is repeated $10^7$ times to form a probability distribution function of region size. This methodology also provides an approximate measure of the mean free path of ionizing photons inside ionized regions.} are shown by the distributions (with circles and crosses indicating their means and medians, respectively).

\subsection{Bubble size -- luminosity relation}
\label{sec:Bubble size -- luminosity relation}

In this section we turn our attention to the population of bubbles surrounding a deep selection of galaxies in our simulation. Our main objective here is to investigate the expected connection between the size of such regions (in terms of mean radius, $\bar{R}$) and the luminosity of the brighest galaxy within them (in terms of their intrinsic absolute UV magnitude, $M_{\rm UV}$), aiming to establish a simple relationship between these properties as a function of redshift. We do so anticipating its use in Section~\ref{sec:Detectability}, where we examine the prospects of detecting ionized regions at high redshift. Note that our galaxy number density predictions are based on intrinsic luminosities and do not include dust attenuation as this is not expected to be significant at such high redshifts ($z \geq 9$). This also maintains the good agreement between the \textsc{BlueTides} UVLFs used in this work \citep{WATERS2016} and the results of \citet{OESCH2016}.

\begin{figure*}
\includegraphics[width = \textwidth]{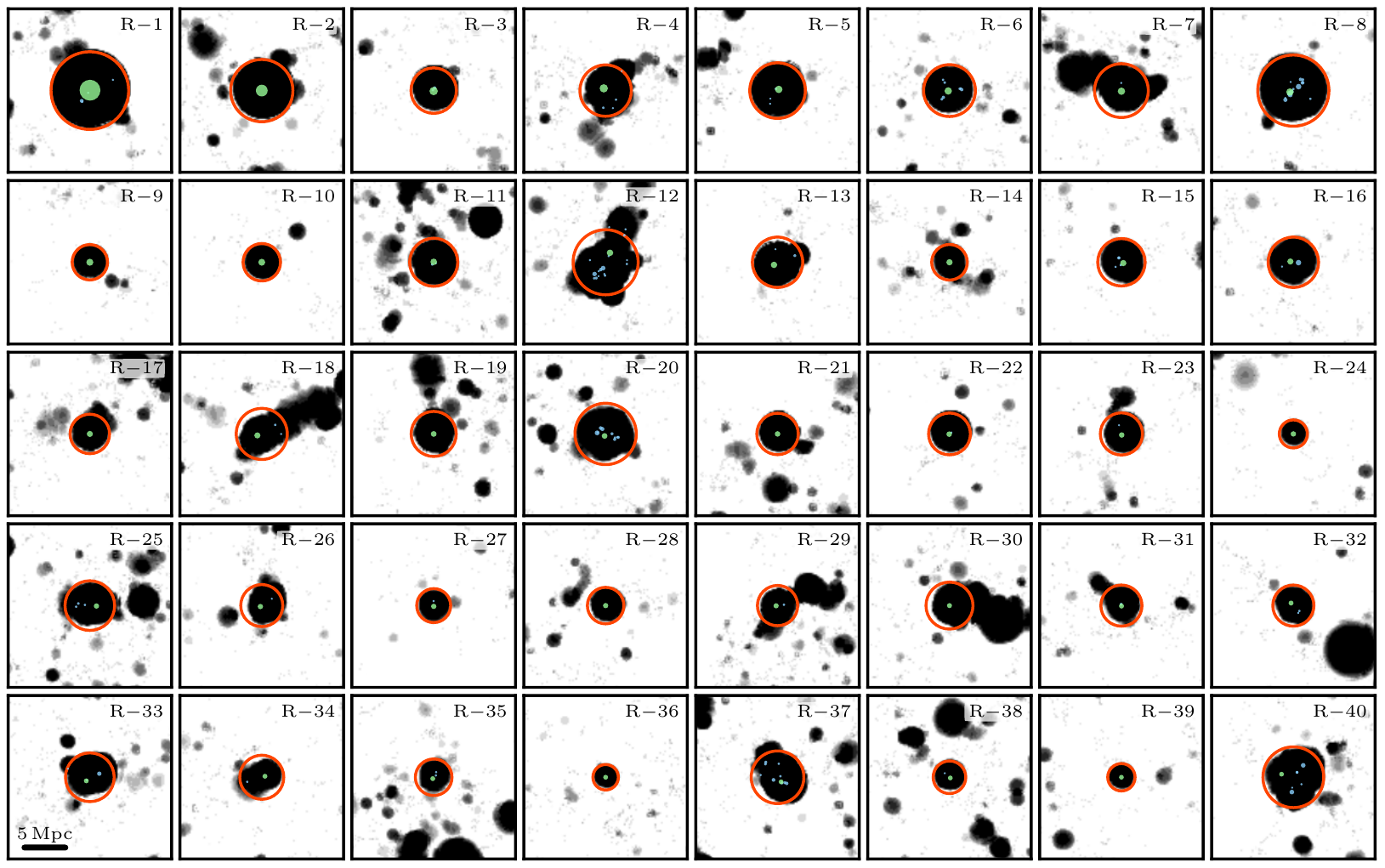}
\caption{Slices through the ionization fields of the first 40 (unique) bubbles surrounding the brightest galaxies at $z = 11.1$. The projected positions of galaxies within the average radius of the bubble (shown by the red circle) are indictated by the filled circles (the brightest in green). Each slice is 1.5\,Mpc deep and has been centred on the centre of luminosity of the galaxies the bubble contains.}
\label{fig:F3}
\end{figure*}

Before demonstrating our $\bar{R}$--$M_{\rm UV}$ fitting procedure we show a sample of zoomed-in slices through the ionization fields of the first 40 (unique) bubbles surrounding the brightest galaxies at $z = 11.1$ in Figure~\ref{fig:F3}. Note that for the purpose of detecting the EoR by stacking bubble 21-cm spectra, we are interested in the relationship between bubble size and the luminosity of the \textit{brightest galaxy in the bubble}. Hence, only one datum contributes to the $\bar{R}$--$M_{\rm UV}$ model fitting for each bubble and therefore the bubbles shown in Figure~\ref{fig:F3} are unique. In this case, since DR-1 and DR-2 are at a significant distance from one another, they happen to be the brightest galaxies in R-1 and R-2, respectively. In general, however, there is no one-to-one correspondence between bubbles and galaxies due to clustering. The average radius of each bubble has been estimated using the individual bubble method described in Section~\ref{sec:GN-z11 analogues: DR-1 and DR-2}. The projected positions of galaxies within the average radius of the bubble (shown by the red circle) are indictated by the filled circles (the brightest in green). Each slice is 1.5\,Mpc deep and, for aesthetics, has been centred on the centre of luminosity of the galaxies the bubble contains. We include this figure in order to demonstrate the variation in geometry of these regions at this redshift.

\begin{figure}
\includegraphics[width = \columnwidth]{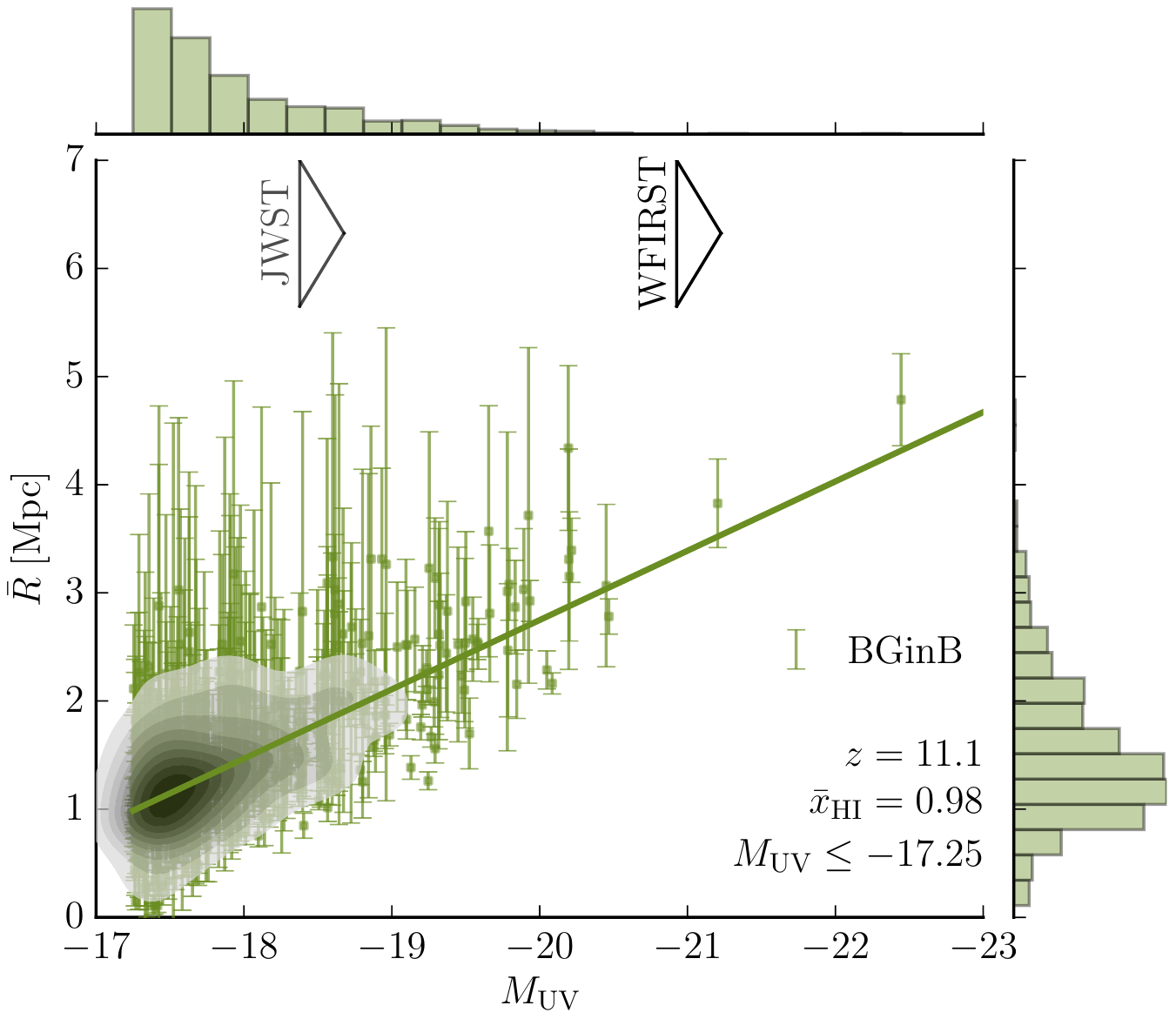}
\caption{Bubble radius versus absolute UV magnitude of the brightest galaxy it contains (labelled `BGinB' for brightest galaxy in bubble) at $z = 11.1$ for $M_{\rm UV} \leq -17.25$. The error bars show the $\pm 1\sigma$ range in radius, calculated using the ray tracing technique described in Section~\ref{sec:GN-z11 analogues: DR-1 and DR-2}. The solid line is the best error-weighted linear fit. The large arrows indicate the {\it WFIRST} 5$\sigma$ and {\it JWST} 8$\sigma$ detection limits at this redshift (see text for details). Histograms on the top and right axes indicate the marginalised distributions of UV magnitude and bubble radius, respectively.}
\label{fig:F4}
\end{figure}

The plot of average bubble radius against the absolute UV magnitude of the brightest galaxy within each bubble (considering galaxies brighter than $M_{\rm UV} = -17.25$ only) at $z = 11.1$ is shown in Figure~\ref{fig:F4}. The error bars represent the $\pm 1\sigma$ range in radius. Histograms on the top and right axes indicate the marginalised distributions of UV magnitude and bubble radius, respectively. For reference, the 5$\sigma$ detection limit of the \emph{WFIRST} HLS \citep[$m_{\rm UV} = 26.75$;][]{Spergel2013} and the 8$\sigma$ detection limit of a wide-field survey using the \emph{James Webb Space Telescope} \citep[\emph{JWST}\footnote{\url{http://www.jwst.nasa.gov/}}, $m_{\rm UV} = 29.3$;][]{Mason2015} at this redshift are indicated. We use this result, and similar results at other redshifts (again, considering galaxies brighter than $M_{\rm UV} = -17.25$ only), to perform an error-weighted Markov Chain Monte Carlo (MCMC) parameter estimation to the linear $\bar{R}$--$M_{\rm UV}$ model, $\bar{R} = a_0 + a_1 M_{\rm UV}$, at 18 redshifts between $z \sim 9$--12. We also fit for an estimate of the variance in $a_0$, $\sigma^2_0$. Specific results for $z = 9.2$, 10.2 and 11.1 ($\bar{x}_{\rm H\textsc{i}} \approx 0.85$, 0.94 and 0.98, respectively) are given in Table~\ref{table:best_fit_results}. While one may expect a {\it non-linear} $\bar{R}$--$M_{\rm UV}$ relationship (e.g. $R \propto L_{\rm UV}^{1/3} \propto 10^{-0.4 M_{\rm UV} / 3}$ for a cosmological Str{\"o}mgren sphere generated by a source of UV luminosity, $L_{\rm UV}$), the many other galaxies in the neighbourhood of the brightest galaxy in each bubble enhance the local ionizing photon budget. Bias and clustering of these sources conspire to complicate this relationship. We choose to fit a linear model for both simplicity and the fact that it describes the luminosity enhancement well (see Appendix~\ref{App:R_MUV_relation}).

\begin{table}
\centering
\ra{1.2}
\begin{tabular}{>{\centering}p{7mm} c c c c}
\toprule
$z$ & $\bar{x}_{\rm H\textsc{i}}$ & $a_1$~[Mpc mag$^{-1}$] & $a_0$~[Mpc] & $\sigma^2_0$~[Mpc$^2$] \\
\midrule
\midrule
9.2 & 0.85 & $-1.28^{+0.03}_{-0.03}$ & $-20.66^{+0.63}_{-0.63}$ & $0.84^{+0.05}_{-0.05}$ \\
10.1 & 0.94 & $-0.82^{+0.02}_{-0.02}$ & $-13.00^{+0.45}_{-0.44}$ & $0.27^{+0.02}_{-0.02}$ \\
11.1 & 0.98 & $-0.64^{+0.02}_{-0.02}$ & $-10.06^{+0.41}_{-0.42}$ & $0.11^{+0.01}_{-0.01}$ \\ \hline
\end{tabular}
\caption{Resulting best-fitting error-weighted $\bar{R}$--$M_{\rm UV}$-model parameter values for selected redshifts/neutral fractions. Confidence limits are determined by the 68 per~cent confidence intervals of the marginalised distributions.}
\label{table:best_fit_results}
\end{table}

Using these results we perform non-linear fits in redshift for $a_1$, $a_0$ and $\sigma^2_0$ (the bespoke functional form and best-fitting parameters are given in Appendix~\ref{App:R_MUV_relation}) to enable us to calculate estimates for $\bar{R}$ and $\sigma_{\bar{R}}$ as a function of magnitude and redshift. These provide the statistical description for the mock bubble size distributions (which we assume to be Gaussian) we utilise in Section~\ref{sec:Stack-averaged 21-cm spectra}.

\section{Detectability}
\label{sec:Detectability}

Having established a functional relationship between the typical size of bubbles as a function of both the luminosity of the brightest galaxy within them and its redshift, we now investigate prospects for their detectability---and therefore direct evidence of cosmic reionization. The two principal competing components we consider to be at play here are the strength of the cosmic 21-cm signal and instrumental noise. We describe the formulation of both in Sections~\ref{sec:Cosmic 21cm signal} and \ref{sec:Instrumental noise}, respectively, before presenting our detection strategy in Section~\ref{sec:Stack-averaged 21-cm spectra}.

\subsection{Cosmic 21-cm signal}
\label{sec:Cosmic 21cm signal}

\begin{figure}
\includegraphics[width = \columnwidth]{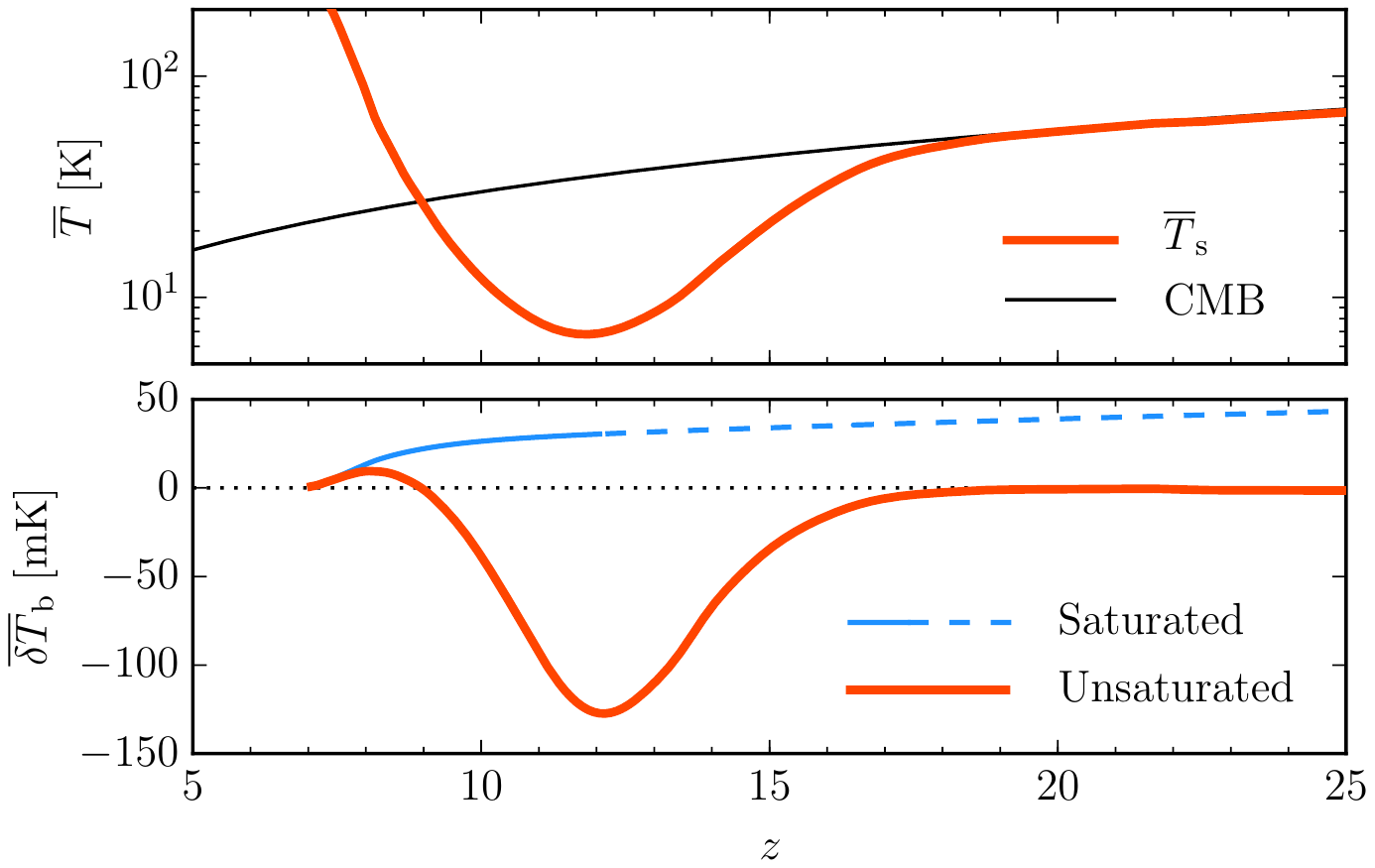}
\caption{Top panel: Evolution of the volume-averaged spin tempertaure according to the Evolution Of 21\,cm Structure simulation \citep{Mesinger2016}. The evolution of the CMB temperature is shown for comparison. Bottom panel: Corresponding volume-averaged 21-cm differential brightness temperature for both the saturated and unsaturated signal case. The dashed extension of the saturated case beyond $z \approx 12$ indicates where the IGM is unlikely to have been fully heated.}
\label{fig:F5}
\end{figure}

The relevant cosmic signal is the spatially-dependent 21-cm differential brightness temperature, $\delta T_{\rm b}$, between hydrogen gas and the CMB along the line of sight \citep[for a detailed discussion of the fundamental physics of the 21-cm line, see, e.g.,][]{FOB2006}. For $z \gg 1$, $\delta T_{\rm b}$ can be written as
\begin{equation}
\begin{split}
\delta T_{\rm b} &\approx 27 x_{\rm H\textsc{i}} (1 + \delta) \left( 1 - \frac{T_\gamma}{T_{\rm s}} \right) \left( \frac{1 + z}{10} \right)^{1/2} \left( \frac{0.15}{\Omega_{\rm m}h^2} \right)^{1/2}\\
&\times \left( \frac{\Omega_{\rm b}h^2}{0.023} \right)\ {\rm mK},
\end{split}
\label{eq:deltaTb}
\end{equation}
where $\delta = \delta(\mathbfit{x}, z) \equiv \rho(\mathbfit{x}, z)/\bar{\rho}(z) - 1$ is the local dark matter overdensity at position $\mathbfit{x}$ and redshift $z$, $x_{\rm H\textsc{i}} = x_{\rm H\textsc{i}}(\mathbfit{x}, z)$ the local neutral fraction, $T_{\rm s} = T_{\rm s}(\mathbfit{x}, z)$ the local spin temperature, and $T_\gamma = 2.73 (1 + z)$\,K the CMB temperature. By using this formulation we ignore redshift-space distortions. When $T_{\rm s} = T_\gamma$ the 21-cm signal from the IGM vanishes. Similarly, as reionization progress ($x_{\rm H\textsc{i}} \rightarrow 0$), the 21-cm signal diminishes. When $T_{\rm s} < T_\gamma$ ($T_{\rm s} > T_\gamma$) the 21-cm signal appears in absorption (emission). In the post-heating regime, where X-rays heat the IGM and the Lyman-$\alpha$ background acts to decouple the 21-cm transition from the CMB (such that $T_{\rm s} \gg T_\gamma$), the 21-cm signal {\it saturates} (since $1 - T_\gamma / T_{\rm s} \rightarrow 1$ in Equation~\ref{eq:deltaTb}) and it appears in emission. This is demonstrated in the top panel of Figure~\ref{fig:F5}, which shows the evolution of the volume-averaged spin tempertaure, $\overline{T}_{\rm s}$, according to the Evolution Of 21\,cm Structure (EOS) simulation by \citet{Mesinger2016} using their `bright galaxies' model. In this model, reionization is dominated by galaxies inside $> 10^{10}$\,M$_\odot$ haloes (roughly corresponding to $M_{\rm UV} \lsim -17$). The bottom panel shows the corresponding volume-averaged 21-cm differential brightness temperature for both the unsaturated and saturated signal case (the dashed extension of the saturated case beyond $z \approx 12$ indicates where the IGM is unlikely to have been fully heated). The EOS simulation incorporates extremely efficient SNe feedback and closely matches the global reionization history of our fiducial model. For computational efficiency, we apply the temperature differential factor in Equation~\ref{eq:deltaTb} homogenously to our $\delta T_{\rm b}$ fields (which have been calculated assuming signal saturation) using the EOS spin temperature model (i.e. we simply adjust the mean temperature).

\subsection{Instrumental noise}
\label{sec:Instrumental noise}

We simulate the instrumental noise of SKA1-LOW based on the imaging performance results provided by \citet{SKAdoc} (hereafter SKA2015). As our detection strategy involves using line-of-sight redshifted 21-cm spectra, the spatial structure of the noise (i.e. in the sky plane) does not concern us. Rather, all that is required is the rms noise, $\sigma_{\rm N}$, for an observation made at frequency $\nu$, with frequency channel width $\Delta \nu$, integrated over time $t_{\rm int}$, and smoothed using a synthesised beam of full width half maximum (FWHM) $\theta_{\rm b}$. More specific details are included in Appendix~\ref{App:noise}.

\subsection{Stack-averaged line-of-sight 21-cm spectra}
\label{sec:Stack-averaged 21-cm spectra}

\subsubsection{Individual verses stack-averaged bubbles}
\label{sec:Individual verses stack-averaged bubbles}

In order to provide a sense of the relative brightness temperatures of signals and noise involved, consider the bubble surrounding DR-1 at $z = 11.1$ ($117.6$\,MHz). According to our simulation the globally-averaged neutral fraction of the surounding IGM is 0.98. Therefore, using Equation~\ref{eq:deltaTb} and assuming saturation, the bubble appears as a near-spherical, $\approx 30$\,mK deep `hole' of radius $\approx 5$\,Mpc. The instrumental noise, smoothed in both the sky plane and frequency space on a scale equal to the radius of the bubble ($\approx 2$\,arcminutes and $\approx 240$\,kHz, respectively; large enough to reduce the instrumental noise without completely smoothing out the bubble\footnote{Note that it is possible to smooth in the sky plane on a scale equal to the diameter of the bubble to reduce noise further with a modest loss of contrast of the bubble. However, due to uncertainty in the grism-determined redshift of the target galaxy and the presence of numerous other bubbles in the surrounding IGM, this tactic does not pay off for imaging {\it individual} bubbles and makes identifying the bubble difficult, if not impossible.}) has an rms of $\sigma_{\rm N} \approx 100$\,mK. Not assuming saturation gives a $\approx 100$\,mK signal. The situation improves at $z = 9.2$, but detection is still unlikely, with a $\approx 23$\,mK saturated signal ($\approx 5$\,mK unsaturated) hidden in noise with $\sigma_{\rm N} \approx 10$\,mK.

Given the unlikely prospect of detecting \emph{individual} bubbles surrounding galaxies, we turn our attention to stack-averaging multiple line-of-sight 21-cm spectra in order to improve the signal-to-noise ratio. As an example, we look at the case using the deep and wide near-infrared High Latitude Survey by \emph{WFIRST} to obtain sky position, redshift and UV~magnitude data of galaxies that lie within the SKA deep integration field\footnote{The centre of the $\sim 2200$\,deg$^2$ footprint of the \emph{WFIRST}-HLS lies at a declination corresponding to zenith at the future site of the Australian infrastructure for the SKA, the Murchison Radio-astronomy Observatory (\url{http://www.atnf.csiro.au/projects/askap/site.html}).}. As {\it WFIRST}'s planned spectroscopic survey will not be sufficiently deep to detect the same, relatively faint, galaxy candidates as identified by the HLS imaging survey, accurate redshift determination will require follow-up grism spectroscopy. Redshifts would be estimated by fitting a spectral energy distribution to the spectra, as was done by \cite{OESCH2016} for GN-z11's spectra from the Wide Field Camera 3 (WFC3) slitless grism on the {\it Hubble Space Telescope} ({\it HST}). Furthermore, without emission lines, these galaxies will need to be Lyman break galaxies (as this break is by far the strongest feature in the spectra). The presence of emission lines would likely improve redshift determination considering the high spectral resolution of {\it WFIRST}'s grism \citep[$R$ = 600;][]{Spergel2013}. At the redshifts investigated in this work, it is reasonable to expect that most of the detectable galaxies will exhibit a Lyman break due to the low ionization fraction of the intervening IGM. We also note that spectroscopic follow-up could also be performed using other instruments, e.g. {\it JWST}. These sky position, redshift and UV~magnitude data are then used to stack-average the line-of-sight redshifted 21-cm spectra centred on each target galaxy. Ideally, this would overlay the bubbles' spectral profiles on top of each other, however, since the redshift upon which each spectrum is centred will be subject to an uncertainty due to the uncertainty in the grism-determined redshift of the target galaxy (which we denote by $\sigma_z$), the bubbles will be scattered along the line-of-sight axis. For the case of GN-z11, \citet{OESCH2016} used {\it HST}-WFC3 grism spectroscopy in combination with photometric data from the Cosmic Assembly Near-infrared Deep Extragalactic Legacy Survey (CANDELS) to place this object at $z = 11.09^{+0.08}_{-0.12}$. Even at half this uncertainty the spatial equivalent of this redshift error at this redshift is approximately twice the size of the bubble surrounding DR-1. Given the design specifications based on the slitless spectroscopic survey capability requirements, \citet{Spergel2013} report that \emph{WFIRST} should be able to determine redshift within $\sigma_z \leq 0.001 (1 + z)$. We make a conservative assumption in this work by first setting our fiducial redshift error to $\sigma_z = 0.05$ for all redshifts investigated, but utilise more optimistic values in Section~\ref{sec:IGM properties} where we explore a method to measure properties of the high-redshift IGM.

\subsubsection{Mock observation sets}
\label{sec:Mock observation sets}

We simulate the expected redshifted 21-cm stacked spectrum using the \emph{WFIRST}-HLS galaxy survey as follows. First, we only stack redshifted 21-cm spectra corresponding to galaxies brighter than some UV~magnitude cutoff, $M_{\rm UV}^{\rm cut}$, and which have a redshift that falls within some range, $\Delta z$, centred on $z_{\rm c}$. Next, we calculate the number of such galaxies in a single SKA1-LOW field using the predicted intrinsic UVLFs provided by \citet{WATERS2016} based on the \textsc{BlueTides} simulation\footnote{We use the results from the \textsc{BlueTides} simulation since its (400 $h^{-1}$~Mpc)$^3$ volume includes more rarer bright galaxies than \emph{Tiamat}, giving more appropriate UVLF fits.} (see Appendix~\ref{App:UVLF}). We find this number to be $> 300$ for the redshifts of interest in this work (see the dashed contour lines in the left-hand and central panels of Figure~\ref{fig:F7}). We randomly sample these UVLFs to obtain a redshift and magnitude for each galaxy. Then, using our $\bar{R}$--$M_{\rm UV}$--$z$ model (and estimate for the variance in $R$, $\sigma_{\bar{R}}^2$) described in Section~\ref{sec:Bubble size -- luminosity relation}, together with our chosen value of $\sigma_z$, we form a randomly sampled mock observation set consisting of the tuple $(z, M_{\rm UV}, R, \Delta d)$ for each galaxy, where $R$ is the bubble radius (sampled from a Gaussian with mean $\bar{R}$ and standard deviation $\sigma_{\bar{R}}$), and $\Delta d$ is the spatial offset of the galaxy from the centre of the spectrum (sampled from a zero-mean Gaussian with standard deviation $\sigma_d = c \sigma_z / H(z)$).

The effective SKA1-LOW field of view has been calculated by applying diffraction theory to a circular aperture and depends on both the observed wavelength, $\lambda$, and station diameter, $D_{\rm s}$. The small angle approximation for this gives
\begin{equation}
\Omega_{\rm SKA} \approx \frac{\pi}{4} \left( \frac{\lambda}{D_{\rm s}} \right)^2 {\rm sr},
\label{eq:R_MUV_parameter_z_model}
\end{equation}
providing $\approx 8$~square degrees at 150~MHz, assuming a station diameter of 35 m. Since we find that the number of spectra required for detection is roughly less than half of what we predict is available the field of view is not an active constraint in this work.

\subsubsection{Other considerations}
\label{sec:Other considerations}

There are three points to consider before moving on:
\begin{enumerate}
\item Despite the high temporal cadence of our simulation, we have a relatively limited number of snapshots between $z \sim ~ 9$--11. Furthermore, since the \emph{Tiamat} volume is much smaller than the SKA survey volume, the number of bubbles around galaxies available to stack at each redshift is in deficit. This is evident in Figure~\ref{fig:F4}, which shows only two galaxies (DR-1 and DR-2) brighter than the \emph{WFIRST} detection limit at $z = 11.1$. Also, as demonstrated in Section~\ref{sec:HII regions surrounding the first galaxies}, the ionized regions surrounding the brightest galaxies are relatively spherical and isolated (i.e. non-overlapping) at the redshifts investigated in this work ($z \gsim ~ 9$). For this reason, we generate synthetic spherical bubbles.
\item In order to beat the instrumental noise down to an acceptable level the number of spectra required to be stacked is $N_{\rm spec} \gsim ~ 50$. Stack-averaging this number of randomly selected $\delta T_{\rm b}$-fields results in a relatively smooth spectrum (i.e. fluctuations from bubbles in the IGM are averaged out). We therefore make the approximation of embedding our synthetic bubbles in a flat IGM whose globally-averaged $\delta T_{\rm b}$ is set by Equation~\ref{eq:deltaTb}.
\item In the large-$N_{\rm spec}$ limit, the stack-averaged signal can be approximated by the convolution of the redshift error probability distribution (mapped onto the space of line-of-sight distance and assumed to be Gaussian) and the spectral profile of a bubble of average size $R$, smoothed in the sky plane on the scale of the assumed SKA1-LOW synthesised beam. Depending on the spatial extent corresponding to $\sigma_z$ ($\sigma_d$) relative to the bubble size, the stack-averaged signal will be quasi-Gaussian (in the case where $\sigma_d > R$), or more closely resemble the individual bubble profile (in the case where $\sigma_d < R$). This is demonstrated in Figure~\ref{fig:F8} (see discussion in Section~\ref{sec:Globally-averaged neutral fraction and bubble size}). We take advantage of this property when calculating ensemble detection statistics in Section~\ref{sec:Ensemble statistics}.
\end{enumerate}

Taking these issues into consideration, we construct synthetic redshifted 21-cm spectra in the following manner: For each galaxy in the mock observation set with associated properties $(z, M_{\rm UV}, R, \Delta d)$, we embed a spherical bubble of radius $R$ in a flat IGM volume with $\delta T_{\rm b}$ set by Equation~\ref{eq:deltaTb} (using $\delta = 0$ and $x_{\rm H\textsc{i}}$ equal to the interpolated value of the globally-averged neutral fraction at $z$ based on our fiducial reionization model). Each volume is centred on $z$ and the bubble is offset from the volume centre by $\Delta d$. The brightness temperature field is binned in frequency space and smoothed in the sky plane using a Gaussian beam with a FWHM equal to the average diameter of the bubbles in the set. The `image' in each channel is zero-meaned as interferometers do not measure the DC (zero spacing) mode. The line-of-sight spectrum through the centre of the bubble is then taken and zero-meaned to simulate removal of spectrally-smooth extragalactic foregrounds (see the discussion Section~\ref{sec:Discussion}). Instrumental noise for each spectrum is simulated by randomly sampling a value for each channel based on $\sigma_{\rm N}(\nu, \Delta \nu, \theta_{\rm b}, t_{\rm int})$ (noise in each channel is assumed to be uncorrelated). We centre the noise realisation for all individual spectra on the central redshift, $z_{\rm c}$\footnote{Doing so overestimates the resulting stack-averaged noise since the redshift distribution of the set of mock galaxies is dominated by lower-redshift galaxies (due to the UVLF from which it is drawn).}. The stack-average of each of these spectral components is then calculated.

\subsubsection{Example realisations}
\label{sec:Example realisations}

\begin{figure*}
\includegraphics[width = \textwidth]{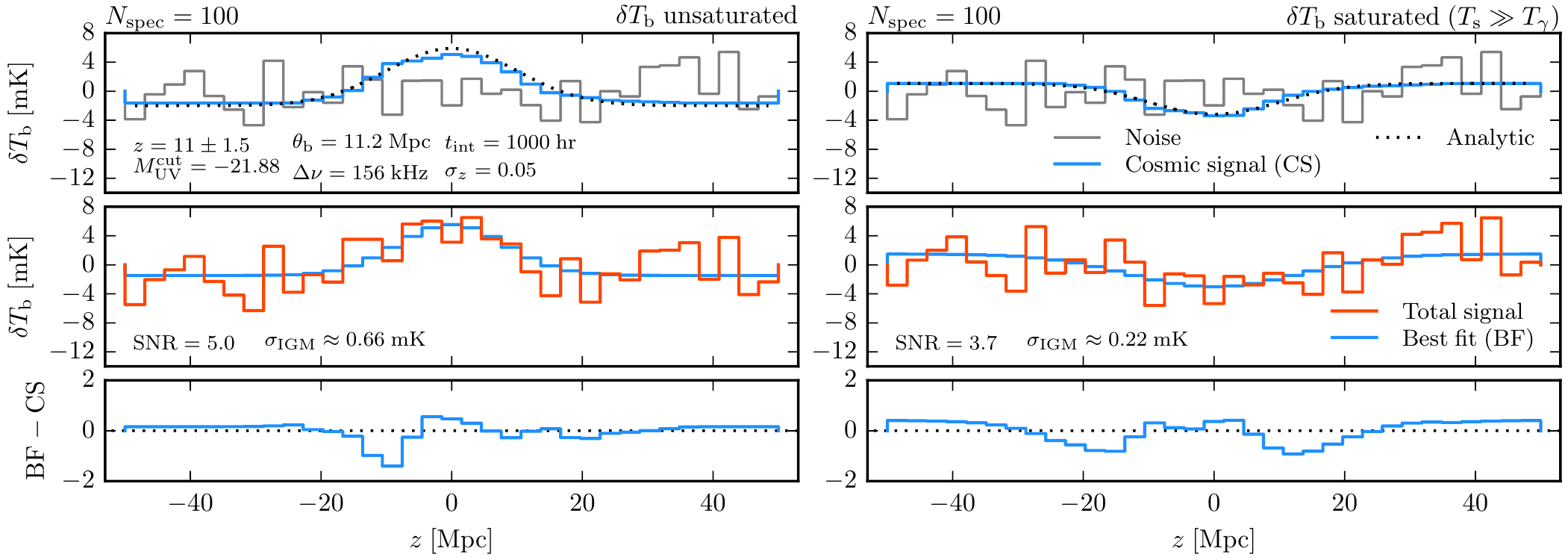}
\caption{Example stack of (100) redshifted 21-cm spectra centred on galaxies brighter than $M_{\rm UV}^{\rm cut} = -21.88$ at $z = 11 \pm 1.5$, assuming an error in grism-determined redshift of $\sigma_z = 0.05$ and a 1000\,hr SKA integration. Smoothing in frequency and the image plane has been performed following the guidelines outlined in Sections~\ref{sec:Other considerations} and \ref{sec:Example realisations}. Left-hand panels show the unsaturated case, while the right-hand panels show the saturated case. Upper panels show the independent components: the cosmic signal (`CS', blue) and instrumental noise (grey). Dotted lines are analytic approximations to the expected stacked cosmic signal. The total signal (cosmic signal $+$ instrumental noise) is shown in the middle panels (red) together with the best-fitting Gaussian model (`BF', blue). The SNR for these realisations are 5.0 and 3.7, while the degree of fluctuation in the surrounding stack-avererged IGM is $\approx 0.66$\,mK and 0.22\,mK ($\ll$~signal) for the unsaturated and saturated cases, respectively. Lower panels show the difference between the best-fitting model and the input cosmic signal.}
\label{fig:F6}
\end{figure*}

Here we demonstrate our spectral stacking strategy with example realisations. Figure~\ref{fig:F6} shows the stack-average of 100 redshifted 21-cm spectra centred on galaxies brighter than $M_{\rm UV}^{\rm cut} = -21.88$ at $z = 11 \pm 1.5$. We assume an error in grism-determined redshifts of $\sigma_z = 0.05$ and a 1000\,hr integration by SKA. The frequency channel width has been set to 156\,kHz (the resulting signal-to-noise ratio is insensitive to the choice of reasonable values of $\Delta \nu$). The left-hand panels show the unsaturated case, while the right-hand panels show the saturated case. The upper panels show the two independent components: the cosmic signal (`CS', blue) and instrumental noise (grey). The dotted lines are the analytic approximations to the expected stacked cosmic signal. The total signal (cosmic signal $+$ instrumental noise) is shown in the middle panels (red) together with the best-fitting Gaussian model (`BF', blue). The Gaussian model is described by two parameters (depth, denoted by $\Delta T$, and standard deviation) and has been zero-meaned and fit with an MCMC parameter estimation technique using flat priors. We calculate the signal-to-noise ratio (SNR) using the resulting marginalised $\Delta T$ distribution, defining it by ${\rm SNR} = \overline{\Delta T} / \sigma_{\Delta T}$, where $\overline{\Delta T}$ is the best-fitting $\Delta T$ value and $\sigma_{\Delta T}$ an estimate of the standard deviation derived from the distribution. The SNR for these example realisations are 5.0 and 3.7 for the unsaturated and saturated cases, respectively. The lower panels show the difference between the input cosmic signal and the best-fitting model (`BF' $-$ `CS'). Any difference is due to a combination of poor sampling (low $N_{\rm spec}$), poor fitting, and non-Gaussianity. The degree of fluctuation in a randomly stack-avererged IGM for these examples is $\sigma_{\rm IGM} \approx 0.66$\,mK and 0.22\,mK (both $\ll \overline{\Delta T}$) for the unsaturated and saturated cases, respectively.

Naturally, the resulting SNR varies with each realisation. For this observation parameter set an ensemble of realisations gives SNRs of $5.6 \pm 0.9$ and $3.1 \pm 0.9$ for the unsaturated and saturated cases, respectively. As expected, stack-averaging a larger number of spectra leads to an improvement, e.g. stacking the brightest 300 ($M_{\rm UV}^{\rm cut} = -21.27$, $z = 11 \pm 1.5$) give SNRs of $7.5 \pm 0.9$ and $4.6 \pm 0.5$ for the unsaturated and saturated cases, respectively. We now go on to explore the full observational parameter space using ensembles of simulations to gauge detectability, both in terms of the expected average and scatter in the SNR.

\subsubsection{Ensemble statistics}
\label{sec:Ensemble statistics}

\begin{figure*}
\includegraphics[width = \textwidth]{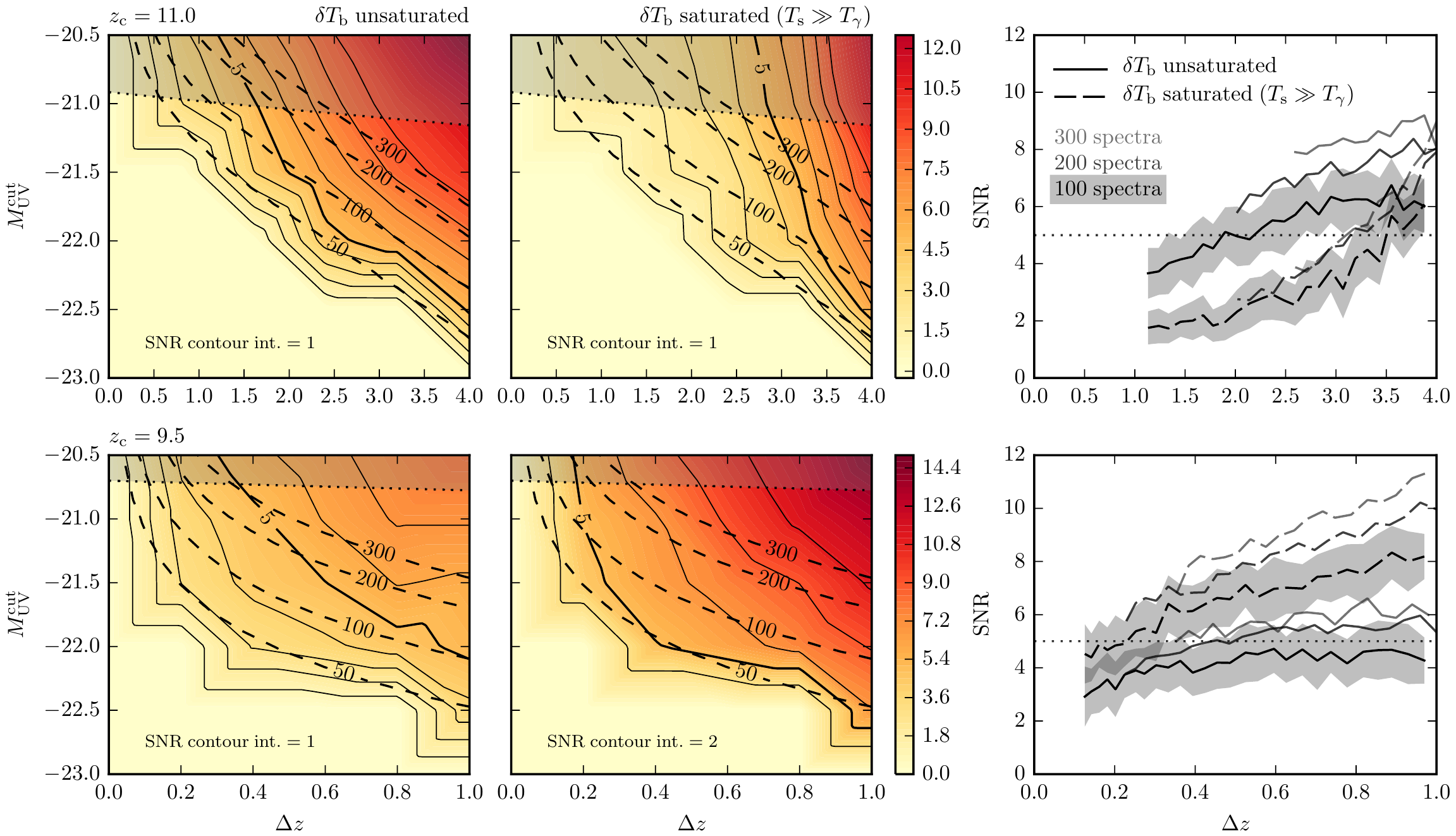}
\caption{Resulting average SNRs for ensembles of realisations exploring the utility of different observation regimes centred on $z_{\rm c} = 11$ (upper panels) and $z_{\rm c} = 9.5$ (lower panels) for the unsaturated case (left panels) and saturated case (middle panels). Dashed contours show the expected number of galaxies in the SKA field of view brighter than $M_{\rm UV}^{\rm cut}$ with redshifts within $z_{\rm c} \pm \Delta z/2$. The colourmaps (and their solid contours) show the estimated average SNR using stacked spectra. The horizontal shaded regions indicate forbidden regimes (dictated by the {\it WFIRST} 5$\sigma$ sensitivity limit). The right-hand panels show the SNR statistics corresponding to the valid sections of the 100-, 200- and 300-spectra contours as a function of $\Delta z$. Solid and dashed lines correspond to the unsaturated and saturated cases, respectively. Shaded regions indicate the 1$\sigma$ confidence regions for the 100-spectra stack results (the variances of the 200- and 300-spectra cases are similar). The dotted line marks a constant SNR of 5.}
\label{fig:F7}
\end{figure*}

The kind of realisations in Section~\ref{sec:Example realisations} can be performed anywhere in the valid ($z_{\rm c},  \Delta z, M_{\rm UV}^{\rm cut}$) observation space. In this section we discuss the average and scatter in SNRs for all the possible observation sets with $z_{\rm c} = 9.5$ and 11, for both the unsaturated and saturated signal case. We calculate these by creating an ensemble of 50 realisations at 100 points in the ($\Delta z, M_{\rm UV}^{\rm cut}$) planes shown in the left-hand and middle column panels of Figure~\ref{fig:F7}. For computational efficiency, we do this using an analytic formulation of the stack-averaged cosmic signal which we have found to converge on realisations with $N_{\rm spec} \gsim ~ 50$. The number of stacked spectra are shown by the dashed line contours, while the colourmaps and unbroken contours show the average SNR (the bold contours mark a constant SNR of 5, which we use as our threshold for detectibility). The relative two-dimensional gradients of these surfaces at these redshifts suggests that the optimal detection strategy (in terms of maximising efficiency) is to stack-average bubbles surrounding the brightest galaxies in the widest redshift range possible. The right-hand panels show the SNR statistics corresponding to the valid sections of the 100-, 200- and 300-spectra contours as a function of $\Delta z$. Shaded regions indicate the 1$\sigma$ confidence regions for the 100-spectra stack results. Only the 100-spectra scatter is shown as the variances of the 200- and 300-spectra cases are similar. The dotted lines mark our constant ${\rm SNR} = 5$ detectibility threshold.

\section{IGM properties}
\label{sec:IGM properties}

\subsection{Spin temperature}
\label{sec:Spin temperature}

The temperature differential-dependent modality of the 21-cm signal (viz. whether it is in emission or absorption) enables a qualitative constraint to be placed on the average spin-temperature of the IGM with respect to the CMB temperature. Given a detection of ionized bubbles using the technique described in the previous section, it is possible to determine if the IGM surrounding them is typically in absorption or emission (see, for example, Figure~\ref{fig:F6}). If the IGM is, on average, in absorption, then $T_{\rm s} < T_\gamma$. On the other hand, if the IGM is, on average, in emission, then $T_{\rm s} > T_\gamma$.

Furthermore, if the signal mode of the IGM was found to change over a range of redshifts this can be used to provide a quantitative measure of the redshift at which $T_{\rm s} = T_\gamma$ and therefore a measure of $T_{\rm s}$ at this redshift. This is not possible for the reionization model presented in this work as the bubbles begin to overlap significantly before the IGM begins to appear in emission. However, this may not be the case in reality as heating by unmodelled sources/mechanisms may occur earlier.

\subsection{Globally-averaged neutral fraction and bubble size}
\label{sec:Globally-averaged neutral fraction and bubble size}

Another IGM property of interest is its globally-averaged neutral fraction, $\bar{x}_{\rm H\textsc{i}}$. Unfortunately, even if an accurate measurement of the differential brightness temperature in Equation~\ref{eq:deltaTb} is made\footnote{This may not be possible using interferometric observations due to the inability to observe the full contrast between fully ionized and not fully-ionized regions (due to instrumental resolution, for example), i.e., no zero-signal baseline can be established.}, this signal depends on both neutral fraction and spin temperature. Therefore, without knowledge of the spin temperature, $\bar{x}_{\rm H\textsc{i}}$ can only be determined when the signal from the IGM is saturated ($1 - T_\gamma / T_{\rm s} \rightarrow 1$). Assuming the IGM is fully heated and the signal appears in saturated emission, we may still be left with a degeneracy in the stacked 21-cm spectra between the average size of the stacked bubbles and the average ionization state of the IGM in which they are embedded. This is because a stack of small bubbles has a similar signature to a stack of larger bubbles in a more ionized IGM (so long as $\sigma_d > R$). This arises due to the grism's limited accuracy and is therefore an observationally-introduced degeneracy, not a physical one. We now demonstrate a method which breaks this degeneracy by taking advantage of the non-Gaussianity and/or constant width of the stacked spectral signal observed where uncertainty in the grism-determined redshifts is small (such that $\sigma_d < R$).

Using $\sigma_z \approx 0.001 (1 + z)$ as {\it WFIRST}'s spectroscopic redshift survey capability \citep{Spergel2013}, we have $\sigma_{z = 10} = 0.011 \equiv 2.4$\,Mpc. Note that this is less than half of the typical radius of bubbles surrounding galaxies detectable by {\it WFIRST} at $z = 10$ according to our simulation. Therefore, rather than being widely spread along the line of sight, the bubbles are relatively tightly aligned on top of each other in the stacked spectrum. As a consequence, their stack-averaged spectrum resembles an instrumentally-smoothed bubble profile rather than a Gaussian with a width equal to that of the grism redshift error distribution. The stacked signal can be seen to be in this regime by inspection (in the case of high signal-to-noise), or by comparing the width of a simple Gaussian fit for the data to $\sigma_d$. Having confirmed $\sigma_d < R$, fitting the analytic model for an instrumentally-smoothed stack of spectra provides estimates for $\bar{R}$ and $\overline{\delta T}_{\rm b}$\footnote{Note that the $\overline{\delta T}_{\rm b}$ estimate is for the average signal of the IGM, not the depth of the trough feature in the observed stacked spectra.}.

\begin{figure}
\includegraphics[width = \columnwidth]{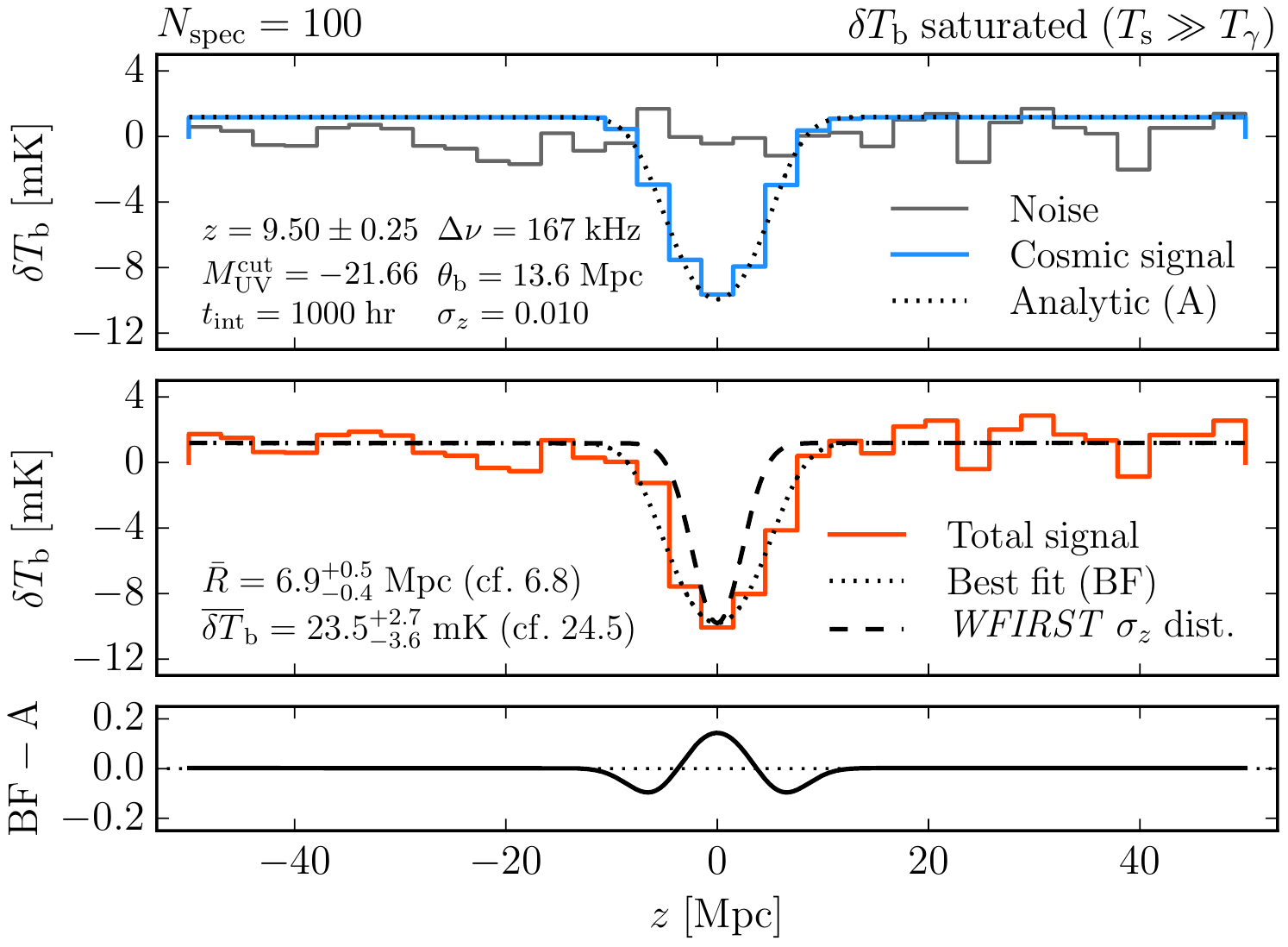}
\caption{Demonstration measuring the average IGM brightness temperature (assumed to be saturated) and bubble size for a mock set of (100) galaxies brighter than $M_{\rm UV}^{\rm cut} = -21.66$ at $z = 9.5 \pm 0.25$. A grism-determined redshift error of $\sigma_z = 0.01$ and a 1000\,hr SKA integration have been used. The upper panel shows the independent signal components and analytic approximation to the expected stacked cosmic signal (`A', dotted line). The total signal is shown in the middle panel together with the best-fitting analytic model (`BF', dotted line) and {\it WFIRST}'s grism redshift error distribution (dashed line). The lower panel shows the difference between the best-fitting model and the analytic model.}
\label{fig:F8}
\end{figure}

This technique is demonstrated for $z_{\rm c} = 9.5$ in Figure~\ref{fig:F8}. The example shown averages 100 redshifted 21-cm spectra centred on galaxies brighter than $M_{\rm UV}^{\rm cut} = -21.66$ with $z = 9.5 \pm 0.25$. We assume an error in grism-determined redshift of $\sigma_z = 0.01$ and a 1000\,hr integration by SKA. The upper panel shows the cosmic signal (blue) and instrumental noise (grey). The dotted line is the analytic approximation to the expected stacked cosmic signal (`A'). The total signal is shown in the middle panel (red) together with its best-fitting analytic model (`BF', dotted line) and {\it WFIRST}'s grism redshift error distribution (dashed line, showing that $\sigma_d < R$). Fitting was performed using an MCMC parameter estimation technique using flat priors with an upper limit on $\overline{\delta T}_{\rm b}$ equal to the value of $\delta T_{\rm b}$ in Equation~\ref{eq:deltaTb} with $\delta = x_{\rm H\textsc{i}} = 1$ and assuming $T_{\rm s} \gg T_\gamma$, ensuring an interpreted estimate for $\bar{x}_{\rm H\textsc{i}}$ that is physically sensible (i.e. $\bar{x}_{\rm H\textsc{i}} \leq 1$). The lower panel shows the difference between the best-fitting model and the analytic model (`BF' $-$ `A'). The resulting best-fitting parameter estimates are $\bar{R} = 6.9^{+0.5}_{-0.4}$\,Mpc (cf. the input value of 6.8\,Mpc) and $\overline{\delta T}_{\rm b} = 23.5^{+2.7}_{-3.6}$\,mK  (cf. the input value of 24.5\,mK). Using Equation~\ref{eq:deltaTb}, we estimate (ignoring any covariance between $\delta T_{\rm b}$ and $z$ in error propagation for simplicity) that $\bar{x}_{\rm H\textsc{i}} = 0.85 \pm 0.13$ (cf. the input value of 0.89).

In the example above, the observed parameter set used was chosen so as to provide a balance between the size of the bubbles stacked (stacking larger bubbles strengthens the signal thereby improving the quality of the fit) and the redshift range of the galaxies, both of which have an impact on the resulting uncertainty in our interpreted estimate for $\bar{x}_{\rm H\textsc{i}}$.

\section{Discussion}
\label{sec:Discussion}

Some sources embedded in an IGM with $T_{\rm s} < T_\gamma$ (i.e. appearing in absorption) will not only ionize their surroundings to form a bubble, but will also heat the gas in their proximity through soft X-ray emission. This will give rise to a relatively thin shell of 21-cm emission beyond the bubble \citep{Tozzi2000, WL2004, Ghara2016}. The $\delta T_{\rm b}$-profile of these sources will therefore resemble a top-hat with `horns'. In order to be conservative we have ignored these effects, although they would improve the signal-to-noise of the stacked spectra used for our detectability predictions in Section~\ref{sec:Detectability}.

Foregrounds were anticipated and have proven to be a significant challenge to detecting the cosmic signal due to their brightness and the chromatic response of the new generation of low-frequency interferometers \citep[see, e.g.,][]{POBER2016}. In this work we have assumed anthropogenic RFI and both Galactic and extragalactic point sources have been removed from the observation data leaving no residual. We have also assumed there are no contamination or removal effects by diffuse Galactic foregrounds (synchrotron emission) apart from the mean removal for each line-of-sight spectra as discussed in Section~\ref{sec:Other considerations}. Previous work has shown that, due to its spectral smoothness, it is possible to subtract this foreground in the imaging regime using a polynomial fitting-based method along each line of sight \citep[see, e.g.,][]{MCQUINN2006, WANG2006, Geil2008b, PandO2011, ALONSO2015}. Since the spectral features we stack in this work are narrow in comparison with the frequency bands over which foreground fitting and subtraction is performed, we expect that these cosmic signatures will not be removed.

\section{Summary and conclusions}
\label{sec:Summary and conclusions}

We have investigated the feasibility of directly detecting regions of ionized hydrogen surrounding galaxies by stacking redshifted 21-cm observations around optically-identified luminous galaxies during early stages of the EoR. In particular, we look at utilising low-frequency observations by the SKA and near infra-red survey data of \emph{WFIRST} in order to image bubbles surrounding massive galaxies at $z \gsim ~ 9.5$.

Our main results can be summarised as follows. We find that:
\begin{enumerate}
\item Our modelling, using the DRAGONS simulation suite, predicts a linear relationship between the size of ionized bubbles and the luminosity of the brighest galaxy within them (in terms of intrinsic absolute UV magnitude, for $M_{\rm UV} \leq \, -17.25$) which evolves with redshift. We provide a fit for this relation and its scatter as a function of redshift.
\item Individual bubbles will not be detected with SKA1-LOW. However, by stacking 100 or more redshifted 21-cm spectra it is possible to detect the EoR directly with a significance of at least 5$\sigma$ at $z \sim 9$--12.
\item Both the spin temperature of the IGM and the accuracy of the grism-determined redshifts of the galaxies have a significant impact on the detectibility of reionization.
\item It is possible to place {\it qualitative} constraints on the evolution of the spin temperature of the IGM at $z \gsim ~ 9$ and it may be possible to {\it quantitatively} measure the redshift at which it is equal to the CMB temperature.
\item Measuring the average size of bubbles and globally-averaged neutral fraction of the IGM is a difficult task due to the degeneracy of these properties' contribution to the cosmic signal. However, if the IGM can be assumed to be fully heated (such that the 21-cm signal is saturated) and the accuracy of the grism-determined redshifts of the galaxies is sufficiently high (compared to the average bubble size of the stack) then this degeneracy may be broken and both the average bubble size and neutral fraction can be accurately determined.
\end{enumerate}

We conclude that imaging 21-cm emission around samples of luminous galaxies from the EoR will provide an additional and complementary probe of cosmic reionization.

\section*{Acknowledgments}

This work was supported by the Victorian Life Sciences Computation Initiative (VLSCI), grant reference UOM0005, on its Peak Computing Facility hosted at The University of Melbourne, an initiative of the Victorian Government. Part of this work was performed on the gSTAR national facility at Swinburne University of Technology. gSTAR is funded by Swinburne University of Technology and the Australian Governments Education Investment Fund. AM acknowledges support from the European Research Council (ERC) under the European Unions Horizon 2020 research and innovation program (grant agreement No 638809 AIDA). This research was funded by the Australian Research Council through the ARC Laureate Fellowship FL110100072 awarded to JSBW. Parts of this research were conducted by the Australian Research Council Centre of Excellence for All-sky Astrophysics (CAASTRO), through project number CE110001020 (PMG). Finally, we would like to thank the anonymous referee for their constructive comments which helped to improve this paper.


\bibliographystyle{mnras}
\bibliography{DRAGONSXII}

\begin{thebibliography}{}
\makeatletter
\relax
\def\mn@urlcharsother{\let\do\@makeother \do\$\do\&\do\#\do\^\do\_\do\%\do\~}
\def\mn@doi{\begingroup\mn@urlcharsother \@ifnextchar [ {\mn@doi@}
  {\mn@doi@[]}}
\def\mn@doi@[#1]#2{\def\@tempa{#1}\ifx\@tempa\@empty \href
  {http://dx.doi.org/#2} {doi:#2}\else \href {http://dx.doi.org/#2} {#1}\fi
  \endgroup}
\def\mn@eprint#1#2{\mn@eprint@#1:#2::\@nil}
\def\mn@eprint@arXiv#1{\href {http://arxiv.org/abs/#1} {{\tt arXiv:#1}}}
\def\mn@eprint@dblp#1{\href {http://dblp.uni-trier.de/rec/bibtex/#1.xml}
  {dblp:#1}}
\def\mn@eprint@#1:#2:#3:#4\@nil{\def\@tempa {#1}\def\@tempb {#2}\def\@tempc
  {#3}\ifx \@tempc \@empty \let \@tempc \@tempb \let \@tempb \@tempa \fi \ifx
  \@tempb \@empty \def\@tempb {arXiv}\fi \@ifundefined
  {mn@eprint@\@tempb}{\@tempb:\@tempc}{\expandafter \expandafter \csname
  mn@eprint@\@tempb\endcsname \expandafter{\@tempc}}}

\bibitem[\protect\citeauthoryear{{Alonso}, {Bull}, {Ferreira}  \&
  {Santos}}{{Alonso} et~al.}{2015}]{ALONSO2015}
{Alonso} D.,  {Bull} P.,  {Ferreira} P.~G.,   {Santos} M.~G.,  2015, \mn@doi
  [MNRAS] {10.1093/mnras/stu2474}, \href
  {http://adsabs.harvard.edu/abs/2015MNRAS.447..400A} {447, 400}

\bibitem[\protect\citeauthoryear{{Bowler} et~al.,}{{Bowler}
  et~al.}{2014}]{Bowler2014}
{Bowler} R.~A.~A.,  et~al., 2014, \mn@doi [MNRAS] {10.1093/mnras/stu449}, \href
  {http://adsabs.harvard.edu/abs/2014MNRAS.440.2810B} {440, 2810}

\bibitem[\protect\citeauthoryear{{Bowman}, {Rogers}  \& {Hewitt}}{{Bowman}
  et~al.}{2008}]{Bowman2008}
{Bowman} J.~D.,  {Rogers} A.~E.~E.,   {Hewitt} J.~N.,  2008, \mn@doi [ApJ]
  {10.1086/528675}, \href {http://adsabs.harvard.edu/abs/2008ApJ...676....1B}
  {676, 1}

\bibitem[\protect\citeauthoryear{{Burns} et~al.,}{{Burns}
  et~al.}{2012}]{DARE2012}
{Burns} J.~O.,  et~al., 2012, \mn@doi [Advances in Space Research]
  {10.1016/j.asr.2011.10.014}, \href
  {http://adsabs.harvard.edu/abs/2012AdSpR..49..433B} {49, 433}

\bibitem[\protect\citeauthoryear{{Cen} \& {Haiman}}{{Cen} \&
  {Haiman}}{2000}]{CH2000}
{Cen} R.,  {Haiman} Z.,  2000, \mn@doi [ApJL] {10.1086/312937}, \href
  {http://adsabs.harvard.edu/abs/2000ApJ...542L..75C} {542, L75}

\bibitem[\protect\citeauthoryear{{Datta}, {Bharadwaj}  \& {Choudhury}}{{Datta}
  et~al.}{2007}]{Datta2007}
{Datta} K.~K.,  {Bharadwaj} S.,   {Choudhury} T.~R.,  2007, \mn@doi [MNRAS]
  {10.1111/j.1365-2966.2007.12421.x}, \href
  {http://adsabs.harvard.edu/abs/2007MNRAS.382..809D} {382, 809}

\bibitem[\protect\citeauthoryear{{Datta}, {Friedrich}, {Mellema}, {Iliev}  \&
  {Shapiro}}{{Datta} et~al.}{2012}]{Datta2012}
{Datta} K.~K.,  {Friedrich} M.~M.,  {Mellema} G.,  {Iliev} I.~T.,   {Shapiro}
  P.~R.,  2012, \mn@doi [MNRAS] {10.1111/j.1365-2966.2012.21268.x}, \href
  {http://adsabs.harvard.edu/abs/2012MNRAS.424..762D} {424, 762}

\bibitem[\protect\citeauthoryear{{Fan}, {Carilli}  \& {Keating}}{{Fan}
  et~al.}{2006}]{Fan2006}
{Fan} X.,  {Carilli} C.~L.,   {Keating} B.,  2006, \mn@doi [ARA\&A]
  {10.1146/annurev.astro.44.051905.092514}, \href
  {http://adsabs.harvard.edu/abs/2006ARA%26A..44..415F} {44, 415}

\bibitem[\protect\citeauthoryear{{Furlanetto}, {Oh}  \& {Briggs}}{{Furlanetto}
  et~al.}{2006}]{FOB2006}
{Furlanetto} S.~R.,  {Oh} S.~P.,   {Briggs} F.~H.,  2006, \mn@doi [Physics
  Reports] {10.1016/j.physrep.2006.08.002}, \href
  {http://adsabs.harvard.edu/abs/2006PhR...433..181F} {433, 181}

\bibitem[\protect\citeauthoryear{{Geil} \& {Wyithe}}{{Geil} \&
  {Wyithe}}{2008}]{Geil2008}
{Geil} P.~M.,  {Wyithe} J.~S.~B.,  2008, \mn@doi [MNRAS]
  {10.1111/j.1365-2966.2008.13159.x}, \href
  {http://adsabs.harvard.edu/abs/2008MNRAS.386.1683G} {386, 1683}

\bibitem[\protect\citeauthoryear{{Geil}, {Wyithe}, {Petrovic}  \& {Oh}}{{Geil}
  et~al.}{2008}]{Geil2008b}
{Geil} P.~M.,  {Wyithe} J.~S.~B.,  {Petrovic} N.,   {Oh} S.~P.,  2008, \mn@doi
  [MNRAS] {10.1111/j.1365-2966.2008.13798.x}, \href
  {http://adsabs.harvard.edu/abs/2008MNRAS.390.1496G} {390, 1496}

\bibitem[\protect\citeauthoryear{{Geil}, {Mutch}, {Poole}, {Angel}, {Duffy},
  {Mesinger}  \& {Wyithe}}{{Geil} et~al.}{2016}]{DRAGONS5}
{Geil} P.~M.,  {Mutch} S.~J.,  {Poole} G.~B.,  {Angel} P.~W.,  {Duffy} A.~R.,
  {Mesinger} A.,   {Wyithe} J.~S.~B.,  2016, \mn@doi [MNRAS]
  {10.1093/mnras/stw1718}, \href
  {http://adsabs.harvard.edu/abs/2016MNRAS.462..804G} {462, 804}

\bibitem[\protect\citeauthoryear{{Ghara}, {Choudhury}  \& {Datta}}{{Ghara}
  et~al.}{2016}]{Ghara2016}
{Ghara} R.,  {Choudhury} T.~R.,   {Datta} K.~K.,  2016, \mn@doi [MNRAS]
  {10.1093/mnras/stw953}, \href
  {http://adsabs.harvard.edu/abs/2016MNRAS.460..827G} {460, 827}

\bibitem[\protect\citeauthoryear{{Ghara}, {Choudhury}, {Datta}  \&
  {Choudhuri}}{{Ghara} et~al.}{2017}]{Ghara2017}
{Ghara} R.,  {Choudhury} T.~R.,  {Datta} K.~K.,   {Choudhuri} S.,  2017,
  \mn@doi [MNRAS] {10.1093/mnras/stw2494}, \href
  {http://adsabs.harvard.edu/abs/2017MNRAS.464.2234G} {464, 2234}

\bibitem[\protect\citeauthoryear{{Greig}, {Mesinger}, {Haiman}  \&
  {Simcoe}}{{Greig} et~al.}{2016}]{Greig2016}
{Greig} B.,  {Mesinger} A.,  {Haiman} Z.,   {Simcoe} R.~A.,  2016, \mn@doi
  [MNRAS] {10.1093/mnras/stw3351}, \href
  {http://adsabs.harvard.edu/abs/2016MNRAS.tmp.1582G} {In press}

\bibitem[\protect\citeauthoryear{{Kohler}, {Gnedin}, {Miralda-Escud{\'e}}  \&
  {Shaver}}{{Kohler} et~al.}{2005}]{Kohler2005}
{Kohler} K.,  {Gnedin} N.~Y.,  {Miralda-Escud{\'e}} J.,   {Shaver} P.~A.,
  2005, \mn@doi [ApJ] {10.1086/444370}, \href
  {http://adsabs.harvard.edu/abs/2005ApJ...633..552K} {633, 552}

\bibitem[\protect\citeauthoryear{{Koopmans} et~al.,}{{Koopmans}
  et~al.}{2015}]{Koopmans2015}
{Koopmans} L.,  et~al., 2015, Advancing Astrophysics with the Square Kilometre
  Array (AASKA14), \href {http://adsabs.harvard.edu/abs/2015aska.confE...1K}
  {p.~1}

\bibitem[\protect\citeauthoryear{{Liu}, {Mutch}, {Angel}, {Duffy}, {Geil},
  {Poole}, {Mesinger}  \& {Wyithe}}{{Liu} et~al.}{2016}]{DRAGONS4}
{Liu} C.,  {Mutch} S.~J.,  {Angel} P.~W.,  {Duffy} A.~R.,  {Geil} P.~M.,
  {Poole} G.~B.,  {Mesinger} A.,   {Wyithe} J.~S.~B.,  2016, \mn@doi [MNRAS]
  {10.1093/mnras/stw1015}, \href
  {http://adsabs.harvard.edu/abs/2016MNRAS.462..235L} {462, 235}

\bibitem[\protect\citeauthoryear{{Majumdar}, {Bharadwaj}  \&
  {Choudhury}}{{Majumdar} et~al.}{2012}]{Majumdar2012}
{Majumdar} S.,  {Bharadwaj} S.,   {Choudhury} T.~R.,  2012, \mn@doi [MNRAS]
  {10.1111/j.1365-2966.2012.21914.x}, \href
  {http://adsabs.harvard.edu/abs/2012MNRAS.426.3178M} {426, 3178}

\bibitem[\protect\citeauthoryear{{Mason}, {Trenti}  \& {Treu}}{{Mason}
  et~al.}{2015}]{Mason2015}
{Mason} C.~A.,  {Trenti} M.,   {Treu} T.,  2015, \mn@doi [ApJ]
  {10.1088/0004-637X/813/1/21}, \href
  {http://adsabs.harvard.edu/abs/2015ApJ...813...21M} {813, 21}

\bibitem[\protect\citeauthoryear{{McQuinn}, {Zahn}, {Zaldarriaga}, {Hernquist}
  \& {Furlanetto}}{{McQuinn} et~al.}{2006}]{MCQUINN2006}
{McQuinn} M.,  {Zahn} O.,  {Zaldarriaga} M.,  {Hernquist} L.,   {Furlanetto}
  S.~R.,  2006, \mn@doi [ApJ] {10.1086/505167}, \href
  {http://adsabs.harvard.edu/abs/2006ApJ...653..815M} {653, 815}

\bibitem[\protect\citeauthoryear{{Mesinger} \& {Furlanetto}}{{Mesinger} \&
  {Furlanetto}}{2007}]{MF2007}
{Mesinger} A.,  {Furlanetto} S.,  2007, \mn@doi [ApJ] {10.1086/521806}, \href
  {http://adsabs.harvard.edu/abs/2007ApJ...669..663M} {669, 663}

\bibitem[\protect\citeauthoryear{{Mesinger}, {Furlanetto}  \& {Cen}}{{Mesinger}
  et~al.}{2011}]{MFC2011}
{Mesinger} A.,  {Furlanetto} S.,   {Cen} R.,  2011, \mn@doi [MNRAS]
  {10.1111/j.1365-2966.2010.17731.x}, \href
  {http://adsabs.harvard.edu/abs/2011MNRAS.411..955M} {411, 955}

\bibitem[\protect\citeauthoryear{{Mesinger}, {Greig}  \& {Sobacchi}}{{Mesinger}
  et~al.}{2016}]{Mesinger2016}
{Mesinger} A.,  {Greig} B.,   {Sobacchi} E.,  2016, \mn@doi [MNRAS]
  {10.1093/mnras/stw831}, \href
  {http://adsabs.harvard.edu/abs/2016MNRAS.459.2342M} {459, 2342}

\bibitem[\protect\citeauthoryear{{Morales} \& {Wyithe}}{{Morales} \&
  {Wyithe}}{2010}]{MW2010}
{Morales} M.~F.,  {Wyithe} J.~S.~B.,  2010, \mn@doi [ARA\&A]
  {10.1146/annurev-astro-081309-130936}, \href
  {http://adsabs.harvard.edu/abs/2010ARA%26A..48..127M} {48, 127}

\bibitem[\protect\citeauthoryear{{Mortlock} et~al.,}{{Mortlock}
  et~al.}{2011}]{Mortlock2011}
{Mortlock} D.~J.,  et~al., 2011, \mn@doi [Nature] {10.1038/nature10159}, \href
  {http://adsabs.harvard.edu/abs/2011Natur.474..616M} {474, 616}

\bibitem[\protect\citeauthoryear{{Mutch}, {Geil}, {Poole}, {Angel}, {Duffy},
  {Mesinger}  \& {Wyithe}}{{Mutch} et~al.}{2016a}]{DRAGONS3}
{Mutch} S.~J.,  {Geil} P.~M.,  {Poole} G.~B.,  {Angel} P.~W.,  {Duffy} A.~R.,
  {Mesinger} A.,   {Wyithe} J.~S.~B.,  2016a, \mn@doi [MNRAS]
  {10.1093/mnras/stw1506}, \href
  {http://adsabs.harvard.edu/abs/2016MNRAS.462..250M} {462, 250}

\bibitem[\protect\citeauthoryear{{Mutch} et~al.,}{{Mutch}
  et~al.}{2016b}]{DRAGONS6}
{Mutch} S.~J.,  et~al., 2016b, \mn@doi [MNRAS] {10.1093/mnras/stw2187}, \href
  {http://adsabs.harvard.edu/abs/2016MNRAS.463.3556M} {463, 3556}

\bibitem[\protect\citeauthoryear{{Oesch} et~al.,}{{Oesch}
  et~al.}{2016}]{OESCH2016}
{Oesch} P.~A.,  et~al., 2016, \mn@doi [ApJ] {10.3847/0004-637X/819/2/129},
  \href {http://adsabs.harvard.edu/abs/2016ApJ...819..129O} {819, 129}

\bibitem[\protect\citeauthoryear{{Oke} \& {Gunn}}{{Oke} \&
  {Gunn}}{1983}]{OG1983}
{Oke} J.~B.,  {Gunn} J.~E.,  1983, \mn@doi [ApJ] {10.1086/160817}, \href
  {http://adsabs.harvard.edu/abs/1983ApJ...266..713O} {266, 713}

\bibitem[\protect\citeauthoryear{{Patra}, {Subrahmanyan}, {Raghunathan}  \&
  {Udaya Shankar}}{{Patra} et~al.}{2013}]{SARAS2013}
{Patra} N.,  {Subrahmanyan} R.,  {Raghunathan} A.,   {Udaya Shankar} N.,  2013,
  \mn@doi [Experimental Astronomy] {10.1007/s10686-013-9336-3}, \href
  {http://adsabs.harvard.edu/abs/2013ExA....36..319P} {36, 319}

\bibitem[\protect\citeauthoryear{{Petrovic} \& {Oh}}{{Petrovic} \&
  {Oh}}{2011}]{PandO2011}
{Petrovic} N.,  {Oh} S.~P.,  2011, \mn@doi [MNRAS]
  {10.1111/j.1365-2966.2011.18276.x}, \href
  {http://adsabs.harvard.edu/abs/2011MNRAS.413.2103P} {413, 2103}

\bibitem[\protect\citeauthoryear{{Planck Collaboration} et~al.,}{{Planck
  Collaboration} et~al.}{2015}]{PLANCK2015}
{Planck Collaboration} et~al., 2015, preprint, \href
  {http://adsabs.harvard.edu/abs/2015arXiv150201589P} {} (\mn@eprint {arXiv}
  {1502.01589})

\bibitem[\protect\citeauthoryear{{Planck Collaboration} et~al.,}{{Planck
  Collaboration} et~al.}{2016}]{PLANCK2016}
{Planck Collaboration} et~al., 2016, \mn@doi [A\&A]
  {10.1051/0004-6361/201628897}, \href
  {http://adsabs.harvard.edu/abs/2016A%26A...596A.108P} {596, A108}

\bibitem[\protect\citeauthoryear{{Pober} et~al.,}{{Pober}
  et~al.}{2016}]{POBER2016}
{Pober} J.~C.,  et~al., 2016, \mn@doi [ApJ] {10.3847/0004-637X/819/1/8}, \href
  {http://adsabs.harvard.edu/abs/2016ApJ...819....8P} {819, 8}

\bibitem[\protect\citeauthoryear{{Poole}, {Angel}, {Mutch}, {Power}, {Duffy},
  {Geil}, {Mesinger}  \& {Wyithe}}{{Poole} et~al.}{2016}]{DRAGONS1}
{Poole} G.~B.,  {Angel} P.~W.,  {Mutch} S.~J.,  {Power} C.,  {Duffy} A.~R.,
  {Geil} P.~M.,  {Mesinger} A.,   {Wyithe} S.~B.,  2016, \mn@doi [MNRAS]
  {10.1093/mnras/stw674}, \href
  {http://adsabs.harvard.edu/abs/2016MNRAS.459.3025P} {459, 3025}

\bibitem[\protect\citeauthoryear{{SKAO Science Team}}{{SKAO Science
  Team}}{2015}]{SKAdoc}
{SKAO Science Team} 2015, SKA1-low Configuration. Document number:
  SKA-SCI-LOW-001. Online; accessed 3-January-2016,
  \url{http://astronomers.skatelescope.org/wp-content/uploads/2015/11/SKA1-Low-Configuration_V4a.pdf}

\bibitem[\protect\citeauthoryear{{Salpeter}}{{Salpeter}}{1955}]{Salpeter1955}
{Salpeter} E.~E.,  1955, \mn@doi [ApJ] {10.1086/145971}, \href
  {http://adsabs.harvard.edu/abs/1955ApJ...121..161S} {121, 161}

\bibitem[\protect\citeauthoryear{{Spergel} et~al.,}{{Spergel}
  et~al.}{2013}]{Spergel2013}
{Spergel} D.,  et~al., 2013, preprint, \href
  {http://adsabs.harvard.edu/abs/2013arXiv1305.5422S} {} (\mn@eprint {arXiv}
  {1305.5422})

\bibitem[\protect\citeauthoryear{{Thompson}, {Moran}  \& {Swenson}}{{Thompson}
  et~al.}{2001}]{TMS}
{Thompson} A.~R.,  {Moran} J.~M.,   {Swenson} Jr. G.~W.,  2001, {Interferometry
  and Synthesis in Radio Astronomy, 2nd Edition}

\bibitem[\protect\citeauthoryear{{Tozzi}, {Madau}, {Meiksin}  \&
  {Rees}}{{Tozzi} et~al.}{2000}]{Tozzi2000}
{Tozzi} P.,  {Madau} P.,  {Meiksin} A.,   {Rees} M.~J.,  2000, \mn@doi [ApJ]
  {10.1086/308196}, \href {http://adsabs.harvard.edu/abs/2000ApJ...528..597T}
  {528, 597}

\bibitem[\protect\citeauthoryear{{Wang}, {Tegmark}, {Santos}  \& {Knox}}{{Wang}
  et~al.}{2006}]{WANG2006}
{Wang} X.,  {Tegmark} M.,  {Santos} M.~G.,   {Knox} L.,  2006, \mn@doi [ApJ]
  {10.1086/506597}, \href {http://adsabs.harvard.edu/abs/2006ApJ...650..529W}
  {650, 529}

\bibitem[\protect\citeauthoryear{{Waters}, {Di Matteo}, {Feng}, {Wilkins}  \&
  {Croft}}{{Waters} et~al.}{2016}]{WATERS2016}
{Waters} D.,  {Di Matteo} T.,  {Feng} Y.,  {Wilkins} S.~M.,   {Croft} R.~A.~C.,
   2016, \mn@doi [MNRAS] {10.1093/mnras/stw2000}, \href
  {http://adsabs.harvard.edu/abs/2016MNRAS.463.3520W} {463, 3520}

\bibitem[\protect\citeauthoryear{{Wyithe} \& {Loeb}}{{Wyithe} \&
  {Loeb}}{2004}]{WL2004}
{Wyithe} J.~S.~B.,  {Loeb} A.,  2004, \mn@doi [ApJ] {10.1086/421042}, \href
  {http://adsabs.harvard.edu/abs/2004ApJ...610..117W} {610, 117}

\bibitem[\protect\citeauthoryear{{Wyithe}, {Loeb}  \& {Barnes}}{{Wyithe}
  et~al.}{2005}]{Wyithe2005}
{Wyithe} J.~S.~B.,  {Loeb} A.,   {Barnes} D.~G.,  2005, \mn@doi [ApJ]
  {10.1086/497160}, \href {http://adsabs.harvard.edu/abs/2005ApJ...634..715W}
  {634, 715}

\makeatother
\end{thebibliography}


\appendix

\section{$\bar{R}$--$M_{\rm UV}$ relation}
\label{App:R_MUV_relation}

\subsection{Assumed linearity}
\label{App:R_MUV_Linearity}

\begin{figure*}
\includegraphics[width = 16cm]{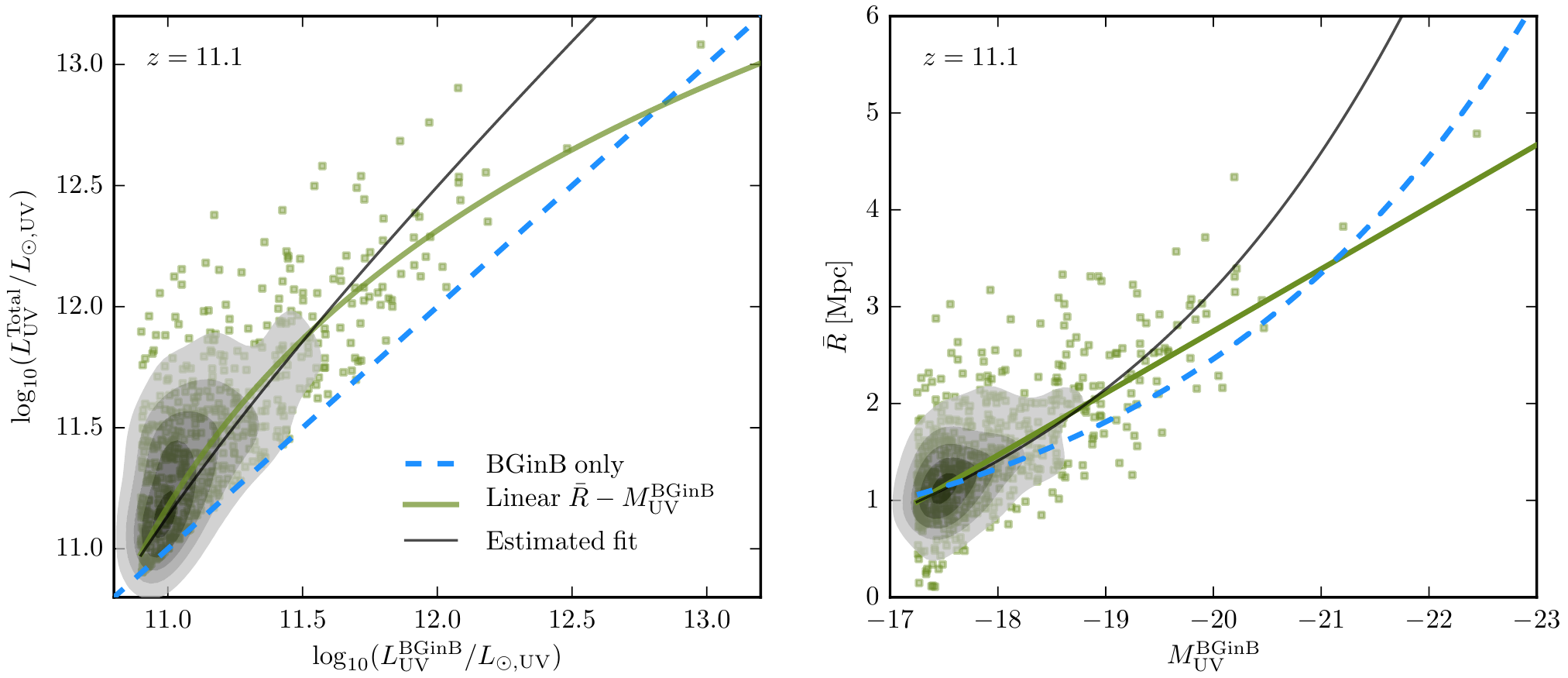}
\caption{Left panel: Total luminosity of galaxies in bubbles as a function of the luminosity of the brightest galaxy in the bubble at $z = 11.1$. Right: Bubble radius as a function of absolute UV magnitude of the brightest galaxy in the bubble. The blue dashed lines correspond to a brightest galaxy-only Str{\"o}mgren sphere relationship, the green lines to the linear $\bar{R}$--$M_{\rm UV}$ relation used in this work, and the thin black lines to an estimate of the $L_{\rm UV}^{\rm Total}$--$L_{\rm UV}^{\rm BGinB}$ relation.}
\label{fig:A0}
\end{figure*}

The typical radius of a cosmological Str{\"o}mgren sphere, $R_{\rm S}$, generated by a source of UV luminosity $L_{\rm UV}$ scales as $R_{\rm S} \propto L_{\rm UV}^{1/3}$ \cite[see, e.g.,][]{CH2000}. In terms of absolute UV magnitude, this gives $R_{\rm S} \propto 10^{-0.4 M_{\rm UV} / 3}$ (i.e. non-linear in $R_{\rm S}$--$M_{\rm UV}$). The $\bar{R}$--$M_{\rm UV}$ plot in Figure~\ref{fig:F4}, however, shows average bubble radius as a function of the absolute UV magnitude of the {\it brightest galaxy only} in each bubble. Of course many other galaxies contribute ionizing photons toward the formation of an ionized region (the brightest of these can be seen in Figures~\ref{fig:F1-nf_seq} and \ref{fig:F3}). These galaxies are both clustered and biased and therefore any enhancement they perform effectively depends on the luminosity of the brightest galaxy in the bubble, $L_{\rm UV}^{\rm BGinB}$, and the size of the region. The total luminosity of galaxies in a region may therefore be written as
\begin{equation}
L_{\rm UV}^{\rm Total} = L_{\rm UV}^{\rm BGinB} + L_{\rm UV}^{\rm NCs}(L_{\rm UV}^{\rm BGinB}, R_{\rm S}),
\label{eq:Ltotal}
\end{equation}
(where NCs stands for `Non-Centrals') and the resulting bubble size scales as $R \propto (L_{\rm UV}^{\rm Total})^{1/3}$. It is this luminosity and scale dependence that alters the simple cube root Str{\"o}mgren sphere relation. This is clearly demonstrated in the left-hand panel of Figure~\ref{fig:A0} which shows $L_{\rm UV}^{\rm Total}$ versus $L_{\rm UV}^{\rm BGinB}$ at $z = 11.1$ (similar results are obtained at other redshifts). In this plot the blue dashed line corresponds to the $L_{\rm UV}^{\rm Total} = L_{\rm UV}^{\rm BGinB}$ (brightest galaxy-only Str{\"o}mgren sphere) case, the green line to the linear $\bar{R}$--$M_{\rm UV}$ relation used in this work (normalised to best fit the data), and the thin black line to an estimate of the $L_{\rm UV}^{\rm Total}$--$L_{\rm UV}^{\rm BGinB}$ relation. The important thing to note here is that the data do not suggest a power law relation between $L_{\rm UV}^{\rm Total}$ and $L_{\rm UV}^{\rm BGinB}$ (and therefore the luminosity enhancement, $L_{\rm UV}^{\rm NCs}$, is indeed $L_{\rm UV}^{\rm BGinB}$-dependent). Hence the cube root-in-$L_{\rm UV}^{\rm BGinB}$ Str{\"o}mgren sphere relation cannot hold. The right-hand panel shows the corresponding $R$--$M_{\rm UV}^{\rm BGinB}$ relations (all normalised to best fit the data). This shows the total luminosity-models (green and black lines) appear to fit the data reasonably well (note, however, that the scatter is relatively large). A reduced chi-squared analysis fails to show that any of these models describe the data significantly better than any other. We leave the analytic solution to the $\bar{R}$--$M_{\rm UV}$ relationship to future work.

\subsection{Fitting $\bar{R}$--$M_{\rm UV}$ model parameters in redshift}
\label{App:R_MUV_parameter_z_fitting}

\begin{figure}
\includegraphics[width = \columnwidth]{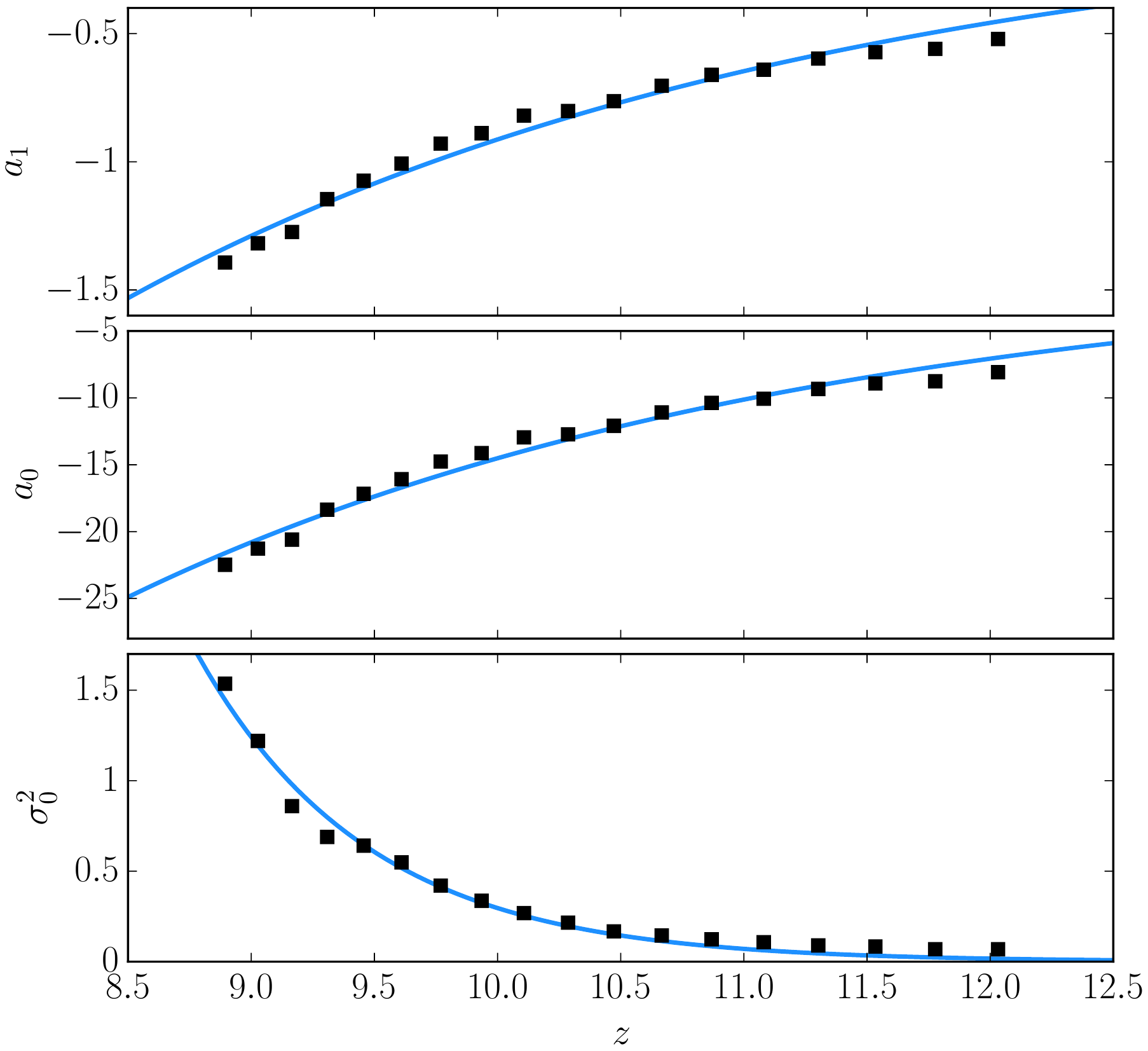}
\caption{Best-fitting error-weighted $\bar{R}$--$M_{\rm UV}$-model parameter values (data points) for eighteen different redshifts between $z \sim 9$--12: $a_1$ (upper panel), $a_0$ (middle panel) and $\sigma^2_0$ (lower panel). Exponential fits are shown by the blue lines.}
\label{fig:A1}
\end{figure}

In Section~\ref{sec:Bubble size -- luminosity relation} we calculated the best-fitting error-weighted $\bar{R}$--$M_{\rm UV}$-model parameter values for eighteen different redshifts between $z \sim 9$--12. These are shown by the data points in Figure~\ref{fig:A1}. To each parameter ($a_1$, $a_0$ and $\sigma^2_0$) we fit the same exponential functional form, given by
\begin{equation}
f(z, c_0, c_1, c_2) = c_0 \exp [ -c_1 (z - c_2) ].
\label{eq:R_MUV_parameter_z_model}
\end{equation}
The resulting best-fitting values for these functional coefficients are given in Table~\ref{table:R_MUV_parameter_z_model_best_fit_results}.

\begin{table}
\centering
\ra{1.2}
\begin{tabular}{@{}lcccc@{}}
\toprule
\multicolumn{2}{c}{$\bar{R}$--$M_{\rm UV}$ model} &  \multicolumn{3}{c}{Function coefficients} \\
Parameter description & Symbol & $c_0$ & $c_1$ & $c_2$ \\
\midrule
\midrule
$\bar{R}$--$M_{\rm UV}$ gradient & $a_1$ & 0.0505 & 0.384 & 17.5 \\
Systematic $\bar{R}$ offset & $a_0$ & 2.32 & 0.401 & 14.5 \\
Variance of systematic offset & $\sigma_0^2$ & 2.73e-3 & 1.37 & 13.4 \\ \hline
\end{tabular}
\caption{Resulting best-fitting values for the functional coefficients of Equation~\ref{eq:R_MUV_parameter_z_model} for each $\bar{R}$--$M_{\rm UV}$ model parameter introduced in Section~\ref{sec:Bubble size -- luminosity relation}.}
\label{table:R_MUV_parameter_z_model_best_fit_results}
\end{table}

\section{SKA1-LOW noise}
\label{App:noise}

This appendix describes how we simulate the instrumental noise of SKA1-LOW. For a comprehensive and authoritative overview of interferometric techniques for radio astronomy, see \citet{TMS}.

\begin{figure}
\includegraphics[width = \columnwidth]{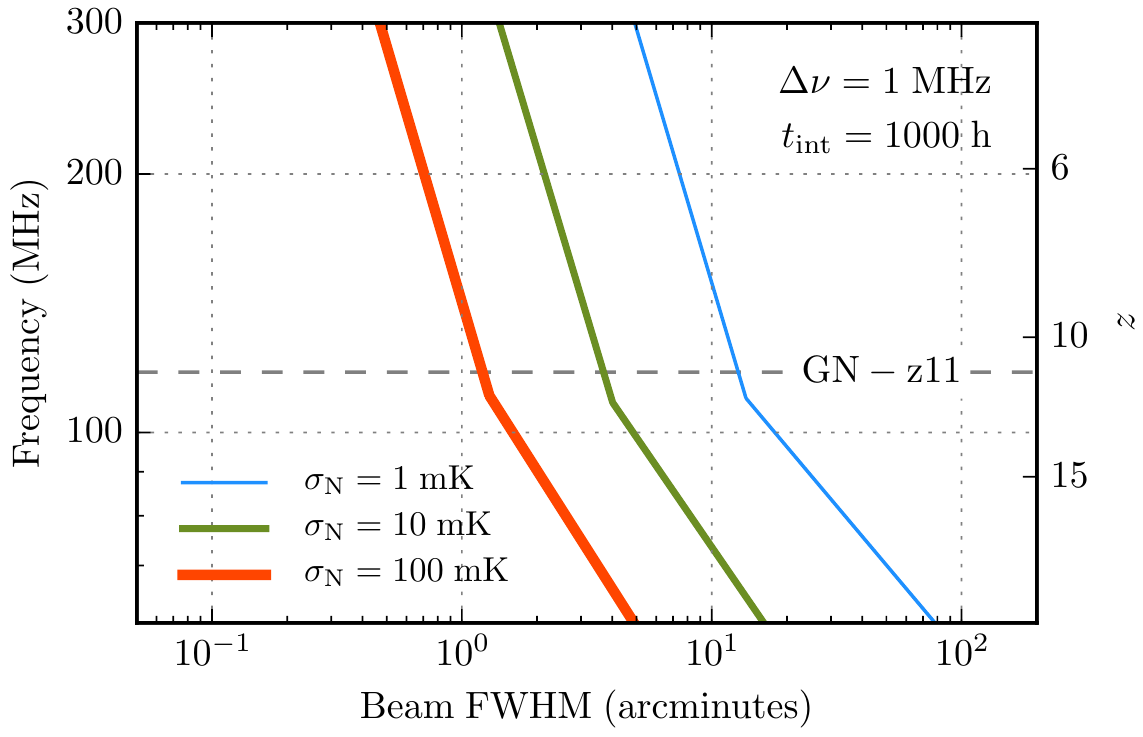}
\caption{Simulated image-space brightness temperature sensitivity for a deep (1000\,hr) integration using a frequency channel width of 1\,MHz as a function of frequency and synthesised beam FWHM (reproduction of Figure~9 in SKA2015). Blue, green and red lines are the 1, 10 and 100\,mK noise isograms, respectively. For reference, the dashed line shows the redshifted 21-cm line frequency corresponding to the redshift of GNz-11.}
\label{fig:B1}
\end{figure}

The image-space noise realisations used throughout this work were generated based on instrumental specifications provided by \citet{SKAdoc} (hereafter SKA2015). In particular, we use their simulated brightness temperature sensitivity results for a deep (1000\,hr) integration as a function of frequency and synthesised beam FWHM (see Figure~9 in SKA2015, or the reproduced version shown in Figure~\ref{fig:B1}). This uses a fiducial frequency channel width of 1\,MHz, however, calculating the sensitivity for different integration times and/or frequency channel widths is possible by noting that
\begin{equation}
\sigma_{\rm N} \propto \frac{1}{\sqrt {\Delta \nu \, t_{\rm int}}}.
\label{eq:rms}
\end{equation}

\section{B{\small{lue}}T{\small{ides}} UVLF}
\label{App:UVLF}

In this work we have used the following double power law to describe the predicted intrinsic UV luminosity functions for galaxies at $z = 9$--11:
\begin{equation}
\phi (M) = \frac{\phi^*}{10^{0.4(\alpha+1)(M-M^*)} + 10^{0.4(\beta+1)(M-M^*)}}
\label{eq:UVLFfunc}
\end{equation}
\citep{Bowler2014}. Here $\phi^*$, $M^*$, $\alpha$ and $\beta$ are the normalisation, characteristic magnitude, faint end slope and bright end slope, respectively. We use the best-fitting values to these parameters, found using the \textsc{BlueTides} simulation by \citet{WATERS2016}. These are:
\begin{equation}
\begin{split}
\ln (\phi^*) &= -[(0.96 \pm 0.22)(1 + z) + (1.57 \pm 2.32)] \\
M^* &= (0.28 \pm 0.12)(1 + z) + (-24.79 \pm 1.41) \nonumber \\
\alpha &= -[(0.14 \pm 0.02)(1 + z) + (0.72 \pm 0.24)] \nonumber \\
\beta &= -[(0.15 \pm 0.05)(1 + z) + (1.78 \pm 0.60)] \nonumber 
\end{split}
\label{eq:UVLFparams}
\end{equation}
where $\phi^*$ is in Mpc$^{-3}$mag$^{-1}$.

\bsp	
\label{lastpage}
\end{document}